\documentclass[a4paper,10pt]{article}
\usepackage{amssymb}
\usepackage{pifont}
\usepackage{graphicx}
\usepackage{setspace}
\usepackage{float}
\providecommand{\keywords}[1]{\textbf{\textit{Index terms---}} #1}
\usepackage{csquotes}
\usepackage{cleveref}
\usepackage{url}
\usepackage{algorithm}
\usepackage[noend]{algpseudocode}
\makeatletter
\def\BState{\State\hskip-\ALG@thistlm}
\makeatother
\usepackage[numbers]{natbib}
\usepackage{usebib}
\newbibfield{author}
\bibpunct{[}{]}{,}{n}{}{,~}
\usepackage{multirow}
\usepackage[table,xcdraw]{xcolor}
\usepackage{lscape}
\usepackage{booktabs}
\usepackage{hhline}
\usepackage{varwidth}
\usepackage{fancyvrb}
\usepackage{tikz}
\usepackage[utf8]{inputenc}
\usepackage{authblk}
\usepackage{xcolor}
\begin{document}
\title {\Huge \textbf{RAMHU: A New Robust Lightweight Scheme for Mutual Users Authentication in Healthcare Applications}}
\author[1,2]{Mishall Al-Zubaidie\thanks{Corresponding author: e-mail: Mishall.Al-Zubaidie@usq.edu.au).}}
\author[2]{Zhongwei Zhang}
\author[2]{Ji Zhang}
\affil[1]{Thi-Qar University, Nasiriya 64001, Iraq}
\affil[2]{Faculty of Health, Engineering and Sciences, University of Southern Queensland, Toowoomba, QLD 4350, Australia}
\date{}

\maketitle
\small  
\noindent{\bf Abstract-} Providing a mechanism to authenticate users in healthcare applications is an essential security requirement to prevent both external and internal attackers from penetrating patients' identities and revealing their health data. Many schemes have been developed to provide authentication mechanisms to ensure that only legitimate users are authorized to connect, but these schemes still suffer from vulnerable security. Various attacks expose patients' data for malicious tampering or destruction. Transferring health-related data and information between users and the health centre makes them exposed to penetration by adversaries as they may move through an insecure channel. In addition, previous mechanisms have suffered from the poor protection of users' authentication information. To ensure the protection of patients' information and data, we propose a scheme that authenticates users based on the information of both the device and the legitimate user. In this paper, we propose a Robust Authentication Model for Healthcare Users (RAMHU) that provides mutual authentication between server and clients. This model utilizes an Elliptic Curve Integrated Encryption Scheme (ECIES) and PHOTON to achieve strong security and a good overall performance. RAMHU relies on multi pseudonyms, physical address, and one-time password mechanisms to authenticate legitimate users. Moreover, extensive informal and formal security analysis with the automated validation of Internet security protocols and applications (AVISPA) tool demonstrates that our model offers a high level of security in repelling a wide variety of possible attacks.\\

\noindent\keywords{Anonymity, ECIES, MAC, mutual authentication, OTP, PHOTON, multi pseudonyms, healthcare applications}\\

\pagenumbering{arabic}
\section{Introduction}
\label{sec:introduction}
\normalsize  
A lack of security and confidentiality of information and data used by healthcare (HC) applications remains the main problem that limits widespread of these applications. These systems require a robust security mechanism to authenticate HC users for achieving the confidentiality, integrity, and availability (CIA) triangle \cite{tp24,tp76} and the compliance with the Health Insurance Portability and Accountability Act (HIPPA) and Health Level Seven (HL7) standards to protect data from being tampered and altered \cite{tp1}. Security requirements (CIA) should be implemented when exchanging data between a client (patient or provider) application and a server application, as any modification to this data affects both medical decisions and the patient's condition \cite{tp9}. Authentication is the first as well as the most critical security requirement that plays a key role in building correct security before the exchange of patients' data in HC applications \cite{tp9,tp15,tp26,tp37}. On one hand, authentication can reduce malicious or fatal errors caused by penetration attacks on the authentication information. On the other hand, it alleviates errors in specifying drug, dose, timing, or procedure \cite{tp2}.  As a result, authentication protocols are a critical requirement to repel various attacks. Typically, the server application should prevent all fake and illegal authentication requests \cite{tp79,tp80}. It should protect personal information, health records, and physiological parameters (such as sugar, and heart rate) \cite{tp9,tp11}. However, authentication information may be easier to compromise if data and information are stored on a single server. Furthermore, the transfer of authentication information in an insecure environment (wireless local area network (WLAN) or Internet), may expose patients' data for detection or modification \cite{tp26,tp27}. Some of the recent cases of security threats on HC applications are presented as follows:
\begin{itemize}
\item Penetration attacks on HC data in the United States' hospitals have occurred (2013). These attacks revealed 85.4\% of the protected health information (PHI) of the 5 largest incidents for patients' records \cite{tp48};
\item Apple Health (Medicaid) was exposed to data breach (2016). This attack revealed 370,000 records of users (Washington state) \cite{tp49};
\item An unauthenticated user penetrated the electronic health record (EHR) containing 14,633 records in the New Jersey Diamond Institute for Fertility and Menopause (2017) \cite{tp13}.
\end{itemize}

The traditional cryptography (such as Rivest Shamir Adleman (RSA)) and signature (such as secure hash algorithm 1 (SHA-1)) schemes require complex computations that consume server resources such as processing power and memory to deal with large amounts of health data in HC applications \cite{tp78} and thus, which could render them unusable. The electronic signature is used to check the integrity of the users' information in the authentication request \cite{tp28}. Many lightweight algorithms such as PHOTON, QUARK, and SPONGENT, are desired to implement electronic signature that perform lightweight operations to reduce high overheads on the server. HC applications require cryptographic and signature high-speed, and secure algorithms \cite{tp3}. To implement an authentication scheme, many algorithms, such as Elliptic Curve Cryptography (ECC), RSA, hash function, bilinear pairing, fuzzy extractor, and XOR operation \cite{tp50}, are used in HC application projects. Many recent HC applications are based on ECC and RSA, both of which provide the same security level, although ECC is more efficient than RSA.The design of an authentication protocols in HC applications should provide mutual authentication, resistance to known attacks such as man-in-the-middle (MITM), eavesdropping, tracing, replay, impersonation, guessing, denial of service (DoS) \cite{tp4}, protection of information and reduced cost and high-efficiency \cite{tp22,tp6}. 

\subsection{Our Contributions}
We propose a Robust Authentication Model for HC Users (RAMHU) for HC applications that perform massive and continuous authentication processes while simultaneously protecting against various attacks. Our contributions include providing robust authentication for legitimate users in the HC applications and access the server repository. They are summarized as follows: 
\begin{itemize}
\item RAMHU uses lightweight algorithms for encryption (ECIES) and signature (PHOTON). These algorithms provide efficient and secure authentication for users in HC applications compared to other algorithms;
\item RAMHU applies a one-time password (OTP) mechanism to authenticate users in their first registration in the HC network with timestamp verification and random nonce generation to repel different types of external attacks; 
\item RAMHU uses a multi pseudonyms mechanism to prevent any association between the real information, pseudonyms, and user's data. This mechanism prevents attackers from identifying HC users (providers and patients); 
\item RAMHU integrates login request with media access control (MAC) address in addition to verifying that this address is original and not fake for authentication of legitimate devices. This prevents attackers from using different devices to compromise the network information;
\item RAMHU improvements the mutual authentication between server and clients to prevent spoofing and impersonation attacks by either fake server or client. This prevents external attacks intended to deceive trusted parties;  
\item We simulate RAMHU with an automated validation of Internet security protocols and applications (AVISPA) tool that is generally acknowledged as an effective way to represent the threat model. We have used AVISPA to prove that our model is secure against both passive and active attacks.
\end{itemize}

\subsection{Structure of the paper}
The remainder of this paper proceeds as follows. Section~\ref{sec:related_work} discusses previous studies related to our research. Threat model and basic concepts about the techniques used in RAMHU will be introduced in Section~\ref{sec:prelininary}. Section~\ref{sec:proposed} describes the proposed authentication model. Section~\ref{sec:security_analysis} describes informal and formal security analysis for our proposed scheme. The conclusion and future research directions are presented in Section~\ref{sec:conclustion}.

\section{Related Work}
\label{sec:related_work}
There are many authentication schemes in literature related to our research area. This section discusses in brief the suggestions in previous studies to design authentication schemes in order to ensure the security of HC users in the network. We investigated these solutions and found that they had different drawbacks. We will discuss the solutions and their disadvantages while clarifying the superiority of our scheme on existing studies.

\citet{tp24} proposed an authentication scheme based on ECC and advanced encryption standards (AES) algorithms. Their scheme uses three entities: user, server, and controller. The authors claimed that their scheme provides many requirements such as mutual authentication, anonymity, and forward confidentiality. The problem with this scheme is that user and controller identities are statically sent to all three entities. If the attacker can penetrate the encryption, he/she can see the related user and controller identities. The attacker can then generate a random number, temporary key, timestamp, and message authentication code. Subsequently, he/she can encrypt the user and controller identities and obtain the message authentication code and send it to the network to become a legitimate and authenticated user. In addition, this scheme uses a 160-bit key with ECC, which is considered unreliable by trusted institutions such as the national institute of standards and technology (NIST). Our scheme uses a 256-bit key and does not exchange real information for legitimate users between clients and servers. An authentication protocol proposed to protect patients' passwords by RSA against off-line password-guessing attacks \cite{tp28}.  The main problem in their scheme is that the authors used RSA with the 1024 key. This algorithm affects the performance of a huge HC network. Many schemes recommend using ECC \cite{tp34,tp41,tp45}, as the ECC-160 is equivalent to RSA-1024 with the same security level. In addition, their protocol suffers from sending an ID clearly from a client to the server at the registration phase, which causes authentication information to be detected for analysis attacks. Using shared-key to implement authentication mechanism is designed to prevent known attacks, especially DoS attacks \cite{tp26}. The authors used the wrong password detection mechanism to reduce the risk of DoS attacks. However, the registration phase of their scheme is not reliable if the ID of patients is sent through an unsafe channel. Additionally, their research suffers from scalability because of the use of the single secret-key mechanism that needs protection from all parties.\\

\citet{tp23} proposed an authentication scheme based on ECC for HC environments. They pointed out that their scheme provides forward secrecy. Their scheme provides authentication during the exchange of messages between the server and RFID (radio-frequency identification)'s tag. However, their scheme shows that the information (identities) and data are stored on a single server, and if the server is hacked, the users' information and data are exposed to detection, tempering, and modification. The ECC and a Petri Nets model are proposed to achieve an authentication requirement to protect HC applications through the mobile cloud \cite{tp16}. The authors pointed out that their scheme is resistant to attacks of eavesdropping, tracking, replay, and spoofing. Unfortunately, their scheme does not address the issue of steal/loss of tag or device and internal attacks that are more serious than external attacks in accessing patients' data as well as no indication of what signature algorithm is used to integrity. In addition, tag's ID is explicitly sent from server to tag, which makes it easier for the attacker to parse the authentication request. \citet{tp14} designed a three-factor (biometric, smart card and password) authentication protocol to protect e-health clouds. Their scheme protects authentication requests from impersonation attacks and off-line password guessing if a mobile device is lost or stolen. Their scheme relies on ECC to support the confidentiality and authentication of HC's users. They used a fuzzy extractor to keep a biometric secret. However, this scheme relies on a single server to authenticate users, which is the target of the attackers. In addition, it performs 7 hash operations that can exhaust the single server capabilities if the network has a huge number of HC users, especially if it is not a lightweight hash. Our model has superior capacity, as it uses separation server (attribute server) for users' information and also only 5 lightweight hash operations in the central server.\\

Cloud-assisted conditional privacy preserving authentication (CACPPA) \cite{tp37} proposed to authenticate the network's nodes. This scheme used ECIES and elliptic curve digital signature algorithm (ECDSA) algorithms with a timestamp and pseudonym integration to perform the authentication process. The problem with this scheme is that it does not provide mutual authentication to prevent an attack from a counterfeit party. Furthermore, authors did not explain the size of the keys in the algorithms to make sure that their scheme was able to repel the various attacks. Furthermore, a single pseudonym cannot separate the link to the real information to prevent analysing and tracking attacks for authentication requests. \citet{tp15} provided a user authentication scheme for HC applications based on AES and hash (SHA-1). They used biometrics users and the anonymity feature to repel attacks such as replay, MITM, and privileged insider. However, their scheme will suffer from key management problem if applied to a large health institution with hundreds of users including managing users' accounts from adding and deleting, as symmetric encryption suffers from a scalability problem, as well as the difficulty of managing the single secret key. Furthermore, the attacker can submit a forgery attack on the login message if it detects the single secret key. Recently, \citet{tp50} provided an authentication scheme based on ECC and hash. Their scheme has supported several servers in user authentication with three factors. In this case, the user can connect to any server to perform the authentication process. In this scheme, the authors did not specify which hash algorithm was used and the size of the message digest (MD), which is necessary for repelling attacks such as collision and preimage. Using multiple servers means that the same users' information is stored on more than one server; thus, the penetration of any server that causes users' information to be detected or modified. Additionally, this scheme did not use a mechanism to prevent the association of real user information with authentication requests such as pseudonyms.\\

\section{The Basic Techniques for our Authentication Scheme}
\label{sec:prelininary}
Authentication mechanism specify connecting users to network services securely. The HC system needs a set of techniques to implement the authentication mechanism before accessing patients' records. One authentication technique will not be sufficient to repel known attacks. To ensure that only legitimate users are associated with the HC application network, our project includes a set of techniques to validate the authentication request. In our scheme, we relied on algorithms that provide lightweight operations and a high-security level for encryption and signature operations. This section describes the threat model and the basic concepts of these techniques.

\subsection{Threat Model}
The threat model has used to detect security weakness in authentication schemes. The threat model is applied to RAMHU based on the Dolev-Yao (dy) \cite{tp52} model in order to build a valid authentication process in insecure environments such as WLAN or Internet. The dy model is an efficient and practical way to illustrate security protocols in the real world. In addition, it is a formal examplar for modelling the risk of attackers against authentication protocols. This model is useful in detecting internal, external, single, and multiple attacks \cite{tp51}. In our threat model, we assume that an intruder is capable of carrying out an internal, external, passive, or active attack such as MITM, replay, eavesdropping, and spoofing. Also, we assume that attributes server ($AS$) is trustworthy. It is safe against repository penetration attacks. We assume threats in our model as follows:
\begin{itemize}
\item The attacker can steal the client application and its files to analyse the data, retrieve the parameters, and reveal the secret key, and then use these applications on different devices.
\item The attacker can listen for authentication requests in the insecure environment and execute interception, replay and  MITM attacks in order to become a legitimate user in the network.
\item  The attacker can execute a forgery or masquerading attack in an attempt to penetrate the authentication process.
\item A legitimate user can perform a privileged insider attack based on his/her legitimacy in the network.
\item  The attacker can successfully guess the real username, password, role and pseudonym associated with it during intensive analysis of many authentication requests.
\end{itemize}

\subsection{Overview of Techniques}
\begin{itemize}
\item Elliptic curve integrated encryption scheme (ECIES) \\
Elliptic Curve Cryptography (ECC) has been used to provide security requirements. It provides confidentiality, integrity, and authentication in the communications network with limited capacity in terms of power and processing. This algorithm was independently proposed by Neal Koblitz and Victor Miller in 1985 \cite{tp81}. It depends on the discrete logarithm problem (DLP), which is impervious to known attacks when selecting parameters accurately \cite{p18}, i.e., difficulty obtaining \textit{k} from \textit{P} and \textit{Q} (where \textit{k} is integer, and \textit{P} and \textit{Q} are two points on the curve). Small parameters used in ECC help to perform computations quickly. These computations are important in constrained-source and large environments that require processing power, memory, or power consumption \cite{tp4,tp82}. ECC provides encryption (Elliptic Curve Integrated Encryption Scheme(ECIES)), signature (Elliptic Curve Digital Signature Algorithm (ECDSA)), and exchange keys (Elliptic Cuvre Diffie-Hellman (ECDH)) approaches \cite{tp81}. Many operations are performed in ECC algorithms (described in four layers) as shown in Figure~\ref{fig1} \cite{p23}. ECIES has provided confidentiality and proven to be extremely efficient in its performance, as it uses small keys; thus, the cost of computation is small compared with other public key cryptography algorithms, such as Rivest Shamir Adleman (RSA) \cite{tp77}. Table~\ref{tab1} \cite{tp82,p99,p100} shows a comparison of key sizes for public key encryption algorithms.
\tikzstyle{block} = [rectangle, draw, fill=white!20, text width=9em, text centered, minimum height=3em ,minimum width=4em]
\tikzstyle{block1} = [rectangle, draw, fill=white!20, text width=5em, text centered, minimum height=3em ,minimum width=4em]
\tikzstyle{block2} = [rectangle, draw, fill=white!20, text width=7em, text centered, minimum height=3em ,minimum width=4em]
\tikzstyle{block3} = [rectangle, draw, fill=white!20, text width=3em, text centered, minimum height=2em ,minimum width=4em]
\tikzstyle{line} = [draw, -latex']   
\tikzstyle{cloud} = [draw, ellipse,fill=white!20, node distance=3cm, minimum height=2em]
\usetikzlibrary{arrows}
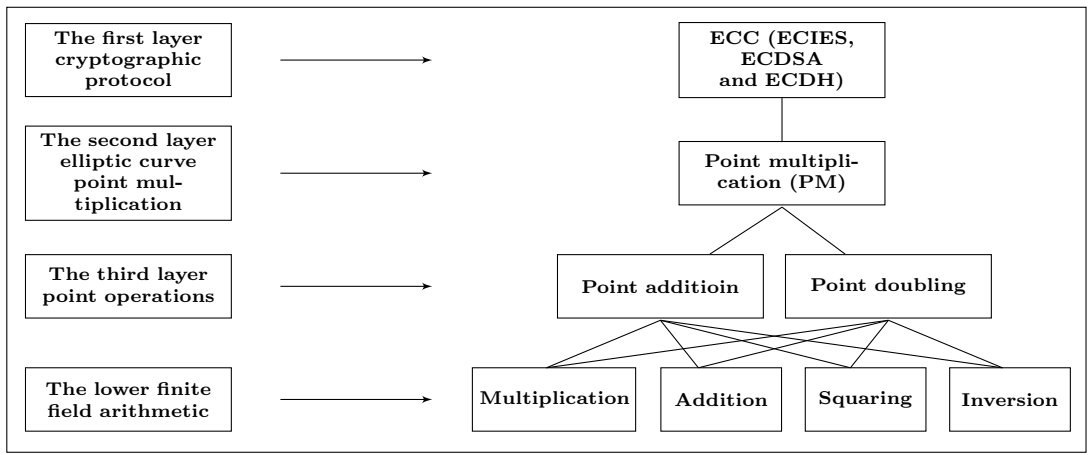
\begin{figure*}[t]
\centering
\begin{tikzpicture}
\scriptsize

\node [block] at (-12,5)(rec1) {\textbf{The first layer \\cryptographic protocol}};
\node [block] at (-12,3.5)(rec2) {\textbf{The second layer\\ elliptic curve point multiplication}};
\node [block] at (-12,2)(rec3) {\textbf{The third layer\\ point operations}};
\node [block] at (-12,0.5)(rec4) {\textbf{The lower finite field arithmetic}};

\node [block] (rec5) at (-3.4,5) {\textbf{ECC (ECIES, ECDSA and ECDH)}};
\node [block] (rec6) at (-3.4,3.5) {\textbf{Point multiplication (PM)}};

\node [block] (rec7) at (-5,2) {\textbf{Point additioin}};
\node [block] (rec8) at (-2,2) {\textbf{Point doubling}};

\node [block2] at (-6.4,0.5)(rec9) {\textbf{Multiplication}};
\node [block1] at (-4.2,0.5)(rec10) {\textbf{Addition}};
\node [block1] at (-2.3,0.5)(rec11) {\textbf{Squaring}};
\node [block1] at (-0.4,0.5)(rec12) {\textbf{Inversion}};

  \path [line] (-10.0,5.0) -- (-8.0,5.0);
  \path [line] (-10.0,3.5) -- (-8.0,3.5);
  \path [line] (-10.0,2.0) -- (-8.0,2.0);
  \path [line] (-10.0,0.5) -- (-8.0,0.5);

 \draw (rec5) --(rec6);
 \draw (-3.4,3.05) --(rec7);
 \draw (-3.4,3.05) --(rec8);
 
 \draw (-5,1.55) --(-6.5,0.92);
 \draw (-5,1.55) --(-4.5,0.92);
 \draw (-5,1.55) --(-2.5,0.92);
 \draw (-5,1.55) --(-0.5,0.92);
 
  \draw (-2.0,1.55) --(-6.5,0.92);
 \draw (-2.0,1.55) --(-4.5,0.92);
 \draw (-2.0,1.55) --(-2.5,0.92);
 \draw (-2.0,1.55) --(-0.5,0.92);

\draw [-triangle 60] (-13.6,5.7) rectangle (0.6,-0.2);
\end{tikzpicture}
 \caption{Arithmetic operations in ECC hierarchy}
  \label{fig1}
\end{figure*}

\begin{table}[!t]
\tiny
\centering
\caption{keys sizes and some information for asymmetric algorithms}
\label{tab1}
\resizebox{\textwidth}{!}{%
\begin{tabular}{|l|l|l|l|l|l|l|l|l|l|}
\hline
\rowcolor[HTML]{EFEFEF} 
Algo. & \multicolumn{5}{l|}{\cellcolor[HTML]{EFEFEF}Keys sizes}                                                                                                                                                                                             & Ratio                                            & Author(s)                                                                  & Year & \begin{tabular}[c]{@{}l@{}}Mathmatical\\ problem\end{tabular}                 \\ \hline
\rowcolor[HTML]{FFFFFF} 
RSA       & & & & & &  & \begin{tabular}[c]{@{}l@{}}Rivest,\\ Shamirand,\\ and Adleman\end{tabular} & 1978 & \begin{tabular}[c]{@{}l@{}}Integer\\ factorization\end{tabular}               \\ \cline{1-1} \cline{8-10} 
Elgamal & \multirow{-2}{*}{1024} & \multirow{-2}{*}{2048} & \multirow{-2}{*}{3072} & \multirow{-2}{*}{7680} & \multirow{-2}{*}{15360} &     & Taher Elgamal & 1985 & \begin{tabular}[c]{@{}l@{}}Multiplicative \\ group\end{tabular} \\ \cline{1-6} \cline{8-10}
ECIES  & 160-223 & 224-255 & 256-383 & 384-511 & 512-more  & \multirow{-3}{*}{1:6-30} & \begin{tabular}[c]{@{}l@{}}Neal Koblitz,\\ and\\ Victor Miller\end{tabular}  & 1985 & \begin{tabular}[c]{@{}l@{}}Elliptic curve\\ discrete log.\\(ECDLP)\end{tabular} \\ \hline
\end{tabular}
}
\end{table}

\item Lightweight hash-function algorithm \\
Hash function has used to generate signatures for authentication requests. PHOTON is lightweight hash-function and extremely suitable for projects that require a robust and reliable signature. This algorithm is based on a sponge-like construction and AES-like primitive for domain extension and permutation efficiency \cite{tp35,tp60,tp57}. PHOTON is available in several versions (80, 128, 160, 224, and 256). It is a balance between the efficiency in the execution of computations on the one hand and security in the implementation of the features of the signature principle that depend on preimage resistant (PR), second preimage resistant (SPR), collision resistant (CR) and mixing-transformation \cite{tp28} on the other hand. It uses sponge construction and applies two phases of absorbing and squeezing to produce message digest (MD); more details are available in \cite{tp53}. Table~\ref{tab2} provides a comparison among the lightweight hash algorithms (the latest version of these algorithms) in terms of security and efficiency \cite{tp60,tp53,tp54,tp55,tp58,tp59,tp61}. Although standard hash algorithms are still used in applications such as SHA-1 (5527 GE, MD 160-bit) and SHA-2 (10868 GE, MD 256-bit), lightweight hash algorithms such as PHOTON (2177 GE, MD 256-bit) provides the most effective solution for handling complex signatures in large HC systems.

Table~\ref{tab2} shows that the PHOTON-256 (2177 GE) provides the best performance compared to all lightweight hash functions and offers a high level of security through the application of signature features PR (224), SPR (128), and CR (128). Linear and differential attacks are the most powerful attacks in the MD analysis of hash functions. Compared to PHOTON, ARMADILLO and SPONGENT-256 also offer signature features, but both are vulnerable to attacks, where ARMIDLLO2 has been attacked by local linearization (practical  semi-free-start collision attack)\cite{tp63} and SPONGENT has been attacked by linear distinguishers  (23 rounds) \cite{tp62} and (13 rounds) \cite{tp56} for all SPONGENT versions; and additionally, they need the most computations (3281 GE and 8653 GE) compared PHOTON-256 (2177 GE). PHOTON is a reliable algorithm against linear and differential attacks \cite{tp53}. It has a high level of security and efficiency; therefore, it is suitable for our project as a signature mechanism.

\begin{table}[!t]
\centering
\tiny
\caption{Comparison of lightweight hash function algorithms}
\label{tab2}
\begin{tabular}{|l|l|l|l|l|l|l|l|l|}
\hline
\rowcolor[HTML]{EFEFEF}  & \multicolumn{2}{l|}{Performace} & & \multicolumn{3}{l|}{Security} & &  \\ \cline{2-3} \cline{5-7}
\rowcolor[HTML]{EFEFEF} 
\multirow{-2}{*}{Algorithm} & \begin{tabular}[c]{@{}l@{}}Gate equivalent\\  area\end{tabular} & \begin{tabular}[c]{@{}l@{}}Throughput\\  (kbps)\end{tabular} & \multirow{-2}{*}{MD} & PR               & SPR              & CR              & \multirow{-2}{*}{Author (s)} & \multirow{-2}{*}{Year} \\ \hline
SQUASH                                              & 2646 GE                                                         & 0.15                                                         & 64                                           & 64               & 64               & 0               & Shamir                                               & 2005                                           \\ \hline
MAME                                                & 8100 GE                                                         & 146.7                                                        & 256                                          & -                & -                & -               & Yoshida et al.                                       & 2007                                           \\ \hline
C-PRESENT-192                                       & 8048 GE                                                         & 59.26                                                        & 192                                          & 192              & 192              & 96              & Bogdanov et al.                                      & 2008                                           \\ \hline
ARMADILLO2                                           & 8653 GE                                                         & 9.38                                                         & 256                                          & 256              & 256              & 128             & Bald et al.                                          & 2010                                           \\ \hline
S-QUARK                                             & 2296 GE                                                         & 3.13                                                         & 256                                          & 224              & 112              & 112             & Aumasson et al.                                      & 2010                                           \\ \hline
KECCAK-f{[}400{]}                                   & 5090 GE                                                         & 14.4                                                         & 128                                          & 128              & 128              & 64              & Kavun and Yalcin                                     & 2010                                           \\ \hline
GLUON                                               & 4724 GE                                                         & 32                                                           & 224                                          & 224              & 112              & 112             & Berger et al.                                        & 2011                                           \\ \hline
SPONGENT-256                                        & 3281 GE                                                         & 11.43                                                        & 256                                          & 240              & 128              & 128             & Bogdanov et al.                                      & 2011                                           \\ \hline
PHOTON-256                                          & 2177 GE                                                         & 0.88                                                         & 256                                          & 224              & 128              & 128             & Guo et al.                                           & 2011                                           \\ \hline
\end{tabular}
\end{table}

\item One time password (OTP)\\
OTP is a way to authenticate legitimate users by generating passcode or nonce only once in a specified time. It will not be applicable the next times for authentication. OTP is an effective method for authenticating users in HC applications if used with robust encryption and signature technologies. Using a static password, or nonce without other authentication mechanisms is weak in respect to attacks. Therefore, OTP provides significant support to the authentication process. This mechanism prevents many attacks such as replay, MITM, and guessing \cite{tp64}. The attacker cannot use this passcode or nonce to connect to the network later. The client sends OTP as part of an authentication request. If the authentication process is valid, the server will delete OTP from the dataset and will not accept it in the future. OTP is a powerful mechanism to mitigate the risk of hackers' communication in the network. In this project, we apply OTP to generate a random password with the first link to users in HC applications to ensure that only legitimate users are connected to the network.

\item Mutual authentication\\
Mutual authentication is a prerequisite for preventing fraudulent authentication requests. The traditional methods of authentication (such as password and name) are not suitable for HC applications \cite{tp3}. In general, there are two kinds of authentication, simple and mutual. In simple authentication, one party performs the authentication process, for example, the server verifies the authentication request for the client. This type of authentication is vulnerable to attacks such as spoofing and impersonation. The attacker can use his device as a fake server to receive all clients' requests. Mutual authentication provides a security solution to prevent known attacks \cite{tp3}. In this type of authentication, each party authenticates the other party and thus, prevents counterfeit attacks by both client or server. In RAMHU, we adopt the mechanism of mutual authentication in the preservation of users' information and data.

\item Media access control (MAC) address\\
MAC address has used to distinguish legitimate users' devices in a network during the completion of the authentication process. All network devices of different types should contain a hardware card (interface) to connect to the local or global network. Each wire or wireless interface has a media access control (MAC) or physical address that consists of a 48-bit. It is divided into six octets and written in hexadecimal such as "8C: 70: 5A: 41: 49: BC". This address is a unique identifier (no duplicate address twice) for a device defined as a global and persistent \cite{tp65}. This address is used in WLAN networks because it offers advantages such as reducing costs and speed in access control procedures \cite{tp66}. It can be changed programmatically in various operating systems such as Linux and Windows. In addition, anyone can use Address Resolution Protocol (ARP) to detect the MAC address of another user in the network \cite{tp67}. The main problem with this address is that the attacker can execute an eavesdropping attack to access the MAC addresses of legitimate devices in the network and then select a legitimate MAC address to use it. For instance, an attacker could execute an ARP poisoning or spoofing attack by using a fake identifier of the MAC address (for a legitimate entity) to gain illegal privileges that would enable it to perform other attacks such as MITM and DOS \cite{tp68}. In addition, randomization operations for the MAC address have become useless in the protection against tracking attacks \cite{tp70,tp71}. Therefore, if the server does not have a mechanism to detect MAC address change, the attacker becomes a legitimate user in the network and has access to network resources.

\end{itemize}

\section{The Proposed Authentication Scheme}
\label{sec:proposed}
In this section, we will detail about RAMHU that provides security and efficiency features in HC applications. This section will be divided into the network model, security goals, and proposed authentication protocols. Notations listed in Table~\ref{tab3} are used to describe symbols used throughout this paper.

\begin{table}[!t]
\begin{center} 
\caption{Paper's notations} 
\label{tab3}
\scriptsize
\setlength{\tabcolsep}{2pt}
\begin{tabular}{|p{95pt}|p{250pt}|}
    \hline
    Symbol   	                         & Description\\ 
    \hline
    $C_{i}$                              & Client entity or user\\ 			
    $CS$                                 & Central server\\			
	$AS$	                             & Attributes server\\
    $DS$                                 & Data server\\ 
    $UID_{i},\ MID_{i}$                            & $C_{i}$'s identity and  medical centre identity\\
    $PW_{i},\ tmpPW_{i}$                 & $C_{i}$'s password, temporary password\\ 
    $C_{i}K_{pu_i},\ C_{i}K_{pr_i}$      & $C_{i}$ public and private key\\
    $CSK_{pu_{i}},\ CSK_{pr_{i}}$        & $CS$ public and private key\\
    $ASK_{pu_{i}},\ ASK_{pr_{i}}$        & $AS$ public and private key\\
    $TS_{C_{i}},\ TS_{CS},\ TS_{AS} $    & Timestamp generated by $C_{i},\ CS,\ AS$ \\    
    $N_{C_{i}}, N_{CS},\ N_{AS}$         & Nonce random generated by $C_{i},\ CS,\ AS$\\ 
    $I$                                  & Internal, or external intruder\\
    $II,\ EI$                                 & Internal intruder and  external intruder\\
    $h(.)$                               & One-way hash function\\
    $R_{i}$                              & Role of patient, patient relative or provider\\
    $RR_{i}$                             & Revocation reason\\    
    $OTP_{i}$                            & One time password to authenticate first time\\
    $C_{i}Sig_{j}$                       & Signature generated by $C_{i}$ and j is signature number\\
    $CSSig_{j}$                          & Signature generated by $CS$\\
    $ASSig_{j}$                          & Signature generated by $AS$\\
    $UP^{CS}_{C_{i}}, PM^{CS}_{C_{i}}$   & User and medical centre pseudonyms sent by $C_{i}$ and verified by $CS$ \\
    $UP^{AS}_{CS}, PM^{AS}_{CS}$         & User and medical centre pseudonyms sent by $CS$ and verified by $AS$\\
    $UP^{CS}_{AS}, PM^{CS}_{AS}$         & User and medical centre pseudonyms sent by $AS$ and verified by $CS$\\
    $UP^{C_{i}}_{CS}, PM^{C_{i}}_{CS}$   & User and medical centre pseudonyms sent by $CS$ and verified by $C_{i}$\\
    $\|,\ \oplus$                                 & Concatenation and exclusive or operations\\  
    $GM,\ CM$                                 & Get MAC and check MAC address\\
    $Enc_{i},\ Dec_{i}$                            & Encryption and decryption operations\\            
    \hline
\end{tabular}
\end{center}
\end{table}

\subsection{Network Model}
\begin{figure}[t]
	\centering
		\includegraphics[width=14cm,height=8cm]{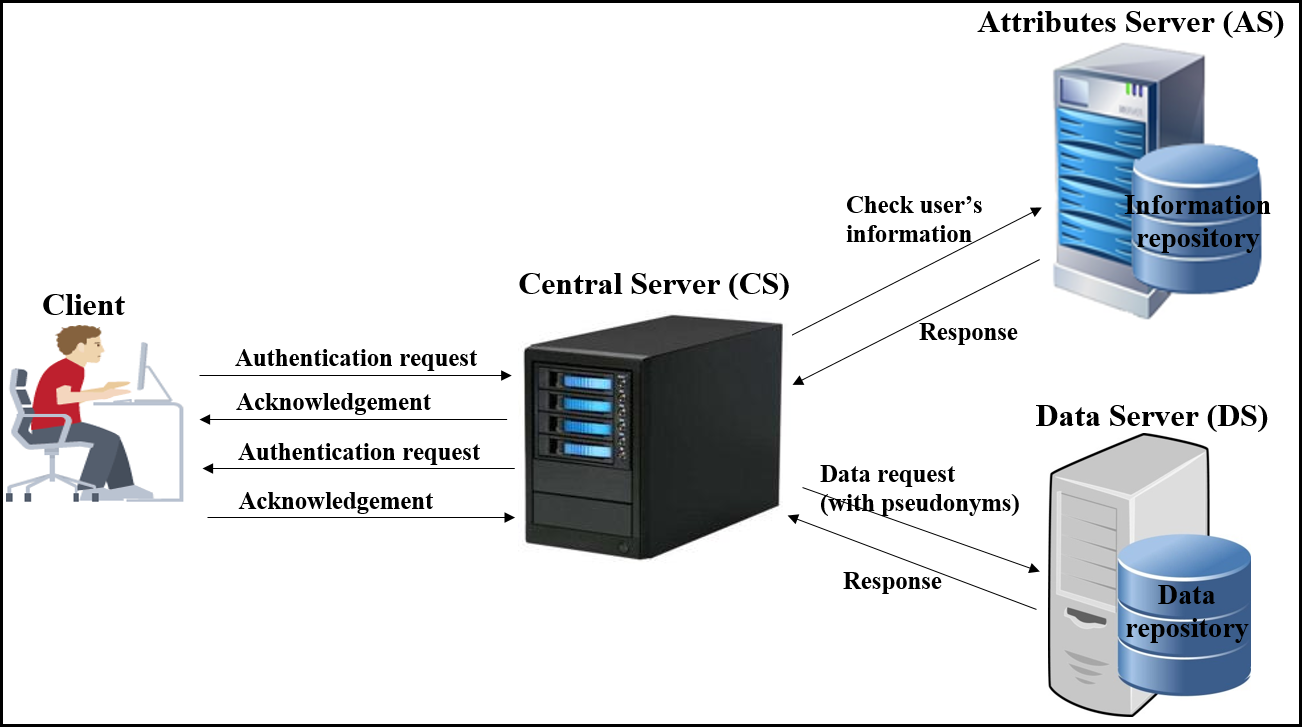}
	\caption{General network model}
	\label{fig3}
\end{figure}

The RAMHU model consists of four entities, as shown in Figure~\ref{fig3}:
\begin{enumerate}
\item  Client ($C_{i}$): This entity includes patients, relatives of patients, and HC providers such as doctors, researchers, emergency practitioners, advisors, and nurses.
\item Central server ($CS$): This entity is a gateway to authenticate users with the attributes server and to authorize the data server.
\item Attributes server ($AS$): This entity contains users' real information as well as multi pseudonyms. The authentication process requires verifying the association of the actual information with the multi pseudonyms in this entity.
\item Data server ($DS$): This entity contains users' data as well as multi pseudonyms. This entity is not implemented in our scheme and is left to future work. Our scheme focuses only on the process of users' authentication in the HC network.
\end{enumerate}

Generally, $C_{i}$ creates an authentication request mainly based on ECIES and PHOTON. $C_{i}$ sends a request to $CS$ to verify encryption, signature, and security parameters (such as MAC address, pseudonyms, and $OTP_i$). Then, the $CS$ sends a request to the $AS$ to verify the link between the pseudonyms, the real information, the signatures, and $PW_i$. After that, the $AS$ sends the response to the $CS$ that verifies the signature and the parameters and then the $CS$ sends the response (authenticated or not) to the $C_{i}$. If the user is authenticated, we assume that the user can then send an authorization request to access the data in the repository ($DS$) and obtain the authorization response from the $CS$, $AS$, and $DS$.

\subsection{Security Goals}
To build a robust authentication scheme, RAMHU adopts the following security requirements:
\begin{itemize}
\item \textbf{Confidentiality}: This requirement is performed to hide authentication information and to preserve user secrecy from detection by intruders. To implement this requirement, a high-security cryptographic algorithm \cite{tp4} should be used. RAMHU executes ECIES to hide authentication information about intruders.

\item \textbf{Integrity}: This is to protect the authentication request information from modification by intruders. The authentication request should reach the intended destination without modification to provide a reliable communication channel for legitimate users \cite{tp38,tp80}. RAMHU performs a PHOTON to prevent any process of modifying the user authentication information in HC applications.

\item \textbf{Non-repudiation}: This requirement prevents both clients and server from denying their authentication requests. This is a way to prove that the message is sent by a particular sender in the HC applications network. If a legitimate user in the network performs internal attacks, he/she cannot deny his/her activities while exploiting the privileges granted to him/her. Our scheme uses PHOTON signatures and a MAC address to meet this requirement and detect malicious attacks.

\item \textbf{Anonymity}: This requirement is extremely important in supporting the confidentiality of the authentication request. The purpose of this requirement is to disguise the source and destination of the authentication request. If the authentication scheme applies anonymity with encryption, the attacker finds it exceedingly difficult to analyze authentication requests for a particular user at different times because the authentication request is different each time the user is connected to the network \cite{tp37}. RAMHU applies this requirement through the use of random nonces between entities.

\item \textbf{Pseudonym}: This requirement denotes the provision of a mechanism to connect non-real attributes (such as terms and symbols) with the real attributes of the user (such as name, address, and phone number). The use of this mechanism in HC applications is an extremely important way to protect the personal information of users and prevent the detection of their identities. RAMHU uses a multi pseudonyms mechanism to prevent and separate the association with real information. 

\item \textbf{Forward secrecy}: This requirement is accomplished when network users use new keys and parameters temporarily without relying on old ones. This requirement prevents attackers from exploiting users' keys and passwords in decrypting authentication requests. Using the temporary random password, private key, and MAC, RAMHU prevents users from accessing previous authentication information.

\item \textbf{Mutual authentication}: This requirement is used in HC applications to mitigate the risks of external fraud. With this feature, each party ensures that it deals with a legitimate party. The server authenticates the client by checking encryptions and signatures and vice versa in order to establish a secure communication channel. In RAMHU, $CS$ and $C_{i}$ authenticate each other to prevent masquerading and impersonating attacks.

\item \textbf{Scalability}: HC applications operate in a scalable environment in terms of data and users. Therefore, these applications require authentication schemes capable of handling and adapting to the ever-increasing number of users of HC applications. This requirement refers to the ability of the authentication scheme to appropriately handle large HC systems. Public key encryption schemes are efficient in supporting this requirement \cite{tp16}.

\item \textbf{Freshness}: This requirement indicates that the authentication request is new and updated to ensure that the attacker cannot replay the authentication request at a later time. This requirement is achieved through the provision of time checking, a random nonce, and change of signatures in each authentication process to counteract counterfeit attacks such as MITM, replay, and impersonation \cite{tp4}, which ensures that the authentication request is unaltered or not tampered with.
\end{itemize}

\subsection{Proposed Authentication Scheme}
RAMHU scheme consists of 5 protocols: initial setup, registration/login, authentication, password update, and revocation. During these protocols, RAMHU provides reliable authentication processes to protect users' information.

\subsubsection{Initial Setup Protocol} 
In this protocol, all entities are ready to start communicating with each other while configuring all security parameters and ECIES's keys as in the following steps:
\begin{itemize}
\item Each legitimate user receives a client application from the authorized system provider.
\item Each legitimate user receives a $PW_i$ (that can be changed later) and a random $OTP_i$ to be used in the first registration.
\item All entities ($C_{i}$, $CS$ and $AS$) should create public and private keys ($C_{i}$ ($CK_{pu_{i}}$, $CK_{pr_{i}}$), $CS$ ($CSK_{pu_{i}}$, $CSK_{pr_{i}}$), and $AS$ ($ASK_{pu_{i}}$, $ASK_{pr_{i}}$)) to be used to validate the authentication request. All entities choose an elliptic curve $Ep (a, b)$ over a prime field $F_{P}$ (where, $P$ = 256) and base point $G$ on curve. Each entity selects a private key $K_{pr_{i}}$ randomly and generates the public key $K_{pu_{i}}$ during the implementation of scalar multiplication ($K_{pu_i} = K_{pr_i} * G$). 
\item All entities broadcast the public key ($CK_{pu_{i}}$, $CSK_{pu_{i}}$, and $ASK_{pu_{i}}$) to use in ECIES's encryption operations.
\end{itemize}

\subsubsection{Registration and Login Protocol} 
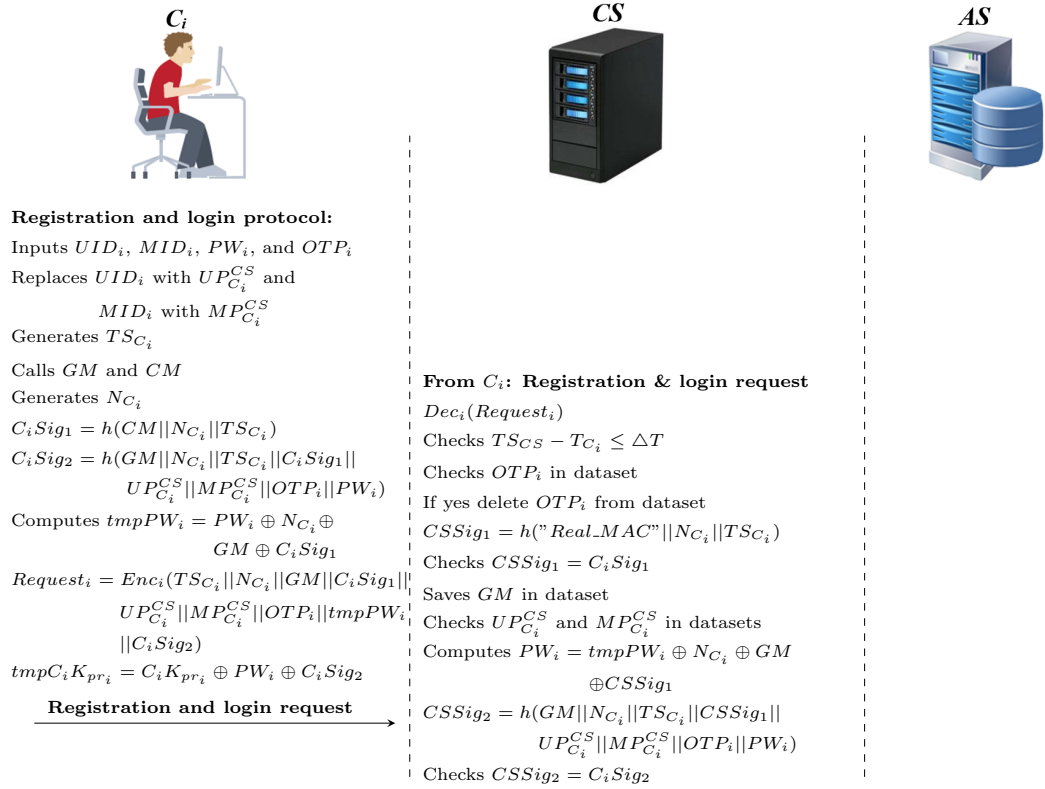
\begin{figure}[t]
\centering
\scriptsize 
\begin{tikzpicture}
\node[inner sep=0pt] (u1) at (2.5,11) {\includegraphics[width=.10\textwidth]{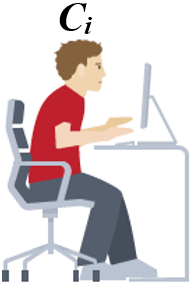}};
\node[inner sep=0pt] (cs1) at (8.0,11) {\includegraphics[width=.10\textwidth]{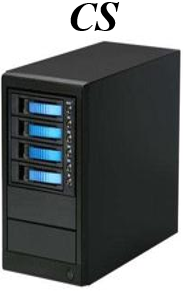}};
\node[inner sep=0pt] (as1) at (13.0,11) {\includegraphics[width=.10\textwidth]{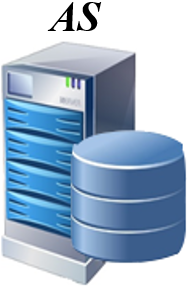}};
\scalebox{0.95}{\scriptsize 
\draw [dashed] (5.7,2.05) -- (5.7,11.0);
\draw [dashed] (12,2.05) -- (12,11.0);
\node [right=-0.8cm] at (25pt,280pt){\textbf{Registration and login protocol:}};
\node [right=-0.8cm] at (25pt,268pt){Inputs $UID_i$, $MID_i$, $PW_i$, and $OTP_i$};
\node [right=-0.8cm] at (25pt,256pt){Replaces $UID_i$ with $UP_{C_i}^{CS}$ and};
\node [right=0.4cm] at (25pt,242.5pt){$MID_i$ with $MP_{C_i}^{CS}$};
\node [right=-0.8cm] at (25pt,232pt) {Generates $TS_{C_i}$};
\node [right=-0.8cm] at (25pt,220pt) {Calls $GM$ and $CM$};
\node [right=-0.8cm] at (25pt,208pt) {Generates $N_{C_i}$};
\node [right=-0.8cm] at (25pt,196pt) {$C_iSig_1 = h(CM||N_{C_i}||TS_{C_i})$};
\node [right=-0.8cm] at (25pt,184pt) {$C_iSig_2= h(GM||N_{C_i}||TS_{C_i}||C_iSig_1||$};
\node [right=0.8cm] at (25pt,172pt) {$UP_{C_i}^{CS}||MP_{C_i}^{CS}||OTP_i||PW_i)$};
\node [right=-0.8cm] at (25pt,159pt) {Computes $tmpPW_i =PW_i\oplus N_{C_i}\oplus$};
\node [right=2.0cm] at (25pt,148pt) {$GM\oplus C_iSig_1$};
\node [right=-0.8cm] at (25pt,136pt) {$Request_i=Enc_i (TS_{C_i}||N_{C_i}||GM||C_iSig_1||$};
\node [right=0.7cm] at (25pt,123pt) {$UP_{C_i}^{CS}||MP_{C_i}^{CS}||OTP_i||tmpPW_i$};
\node [right=0.7cm] at (25pt,111pt) {$||C_iSig_2)$};
\node [right=-0.8cm] at (25pt,99pt) {$tmpC_iK_{pr_i}= C_iK_{pr_i}\oplus PW_i\oplus C_iSig_2$};

\draw [->,>=stealth] (0.5,2.8) -- (5.5,2.8) node[above,pos=0.46] {\textbf{Registration and login request}};
\node [right=4.9cm] at (25pt,215pt){\textbf{From $C_i$: Registration \& login request}};
\node [right=4.9cm] at (25pt,203pt){$Dec_i (Request_i)$};
\node [right=4.9cm] at (25pt,191pt){Checks $TS_{CS}-T_{C_i}\leq \triangle T$};
\node [right=4.9cm] at (25pt,179pt){Checks $OTP_i$ in dataset};
\node [right=4.9cm] at (25pt,167pt){If yes delete $OTP_i$ from dataset};
\node [right=4.9cm] at (25pt,155pt){$CSSig_1 = h("Real\_MAC"||N_{C_i}||TS_{C_i})$};
\node [right=4.9cm] at (25pt,143pt){Checks $CSSig_1= C_iSig_1$};
\node [right=4.9cm] at (25pt,131pt){Saves $GM$ in dataset};
\node [right=4.9cm] at (25pt,119pt){Checks $UP_{C_i}^{CS}$ and $MP_{C_i}^{CS}$ in datasets};
\node [right=4.9cm] at (25pt,107pt){Computes $PW_i =tmpPW_i\oplus N_{C_i}\oplus GM$};
\node [right=7.2cm] at (25pt,95pt){$\oplus CSSig_1$};
\node [right=4.9cm] at (25pt,83pt){$CSSig_2 = h(GM||N_{C_i}||TS_{C_i}||CSSig_1||$};
\node [right=6.5cm] at (25pt,71pt){$UP_{C_i}^{CS}||MP_{C_i}^{CS}||OTP_i||PW_i)$};
\node [right=4.9cm] at (25pt,59pt){Checks $CSSig_2= C_iSig_2$};

} 
\end{tikzpicture}
	\caption{Registration and login protocol}
	\label{fig:registration_and_login_protocol}
\end{figure}

\begin{figure}[t]
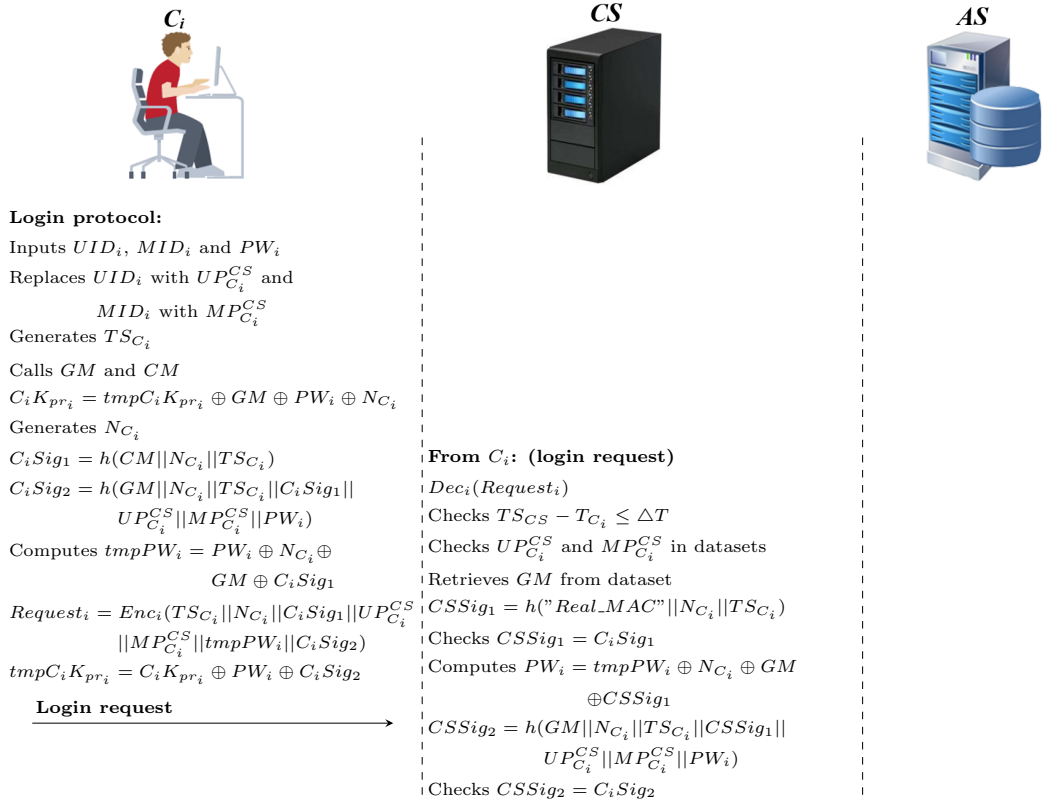

\centering
\scriptsize 
\begin{tikzpicture}
\node[inner sep=0pt] (u1) at (2.5,11) {\includegraphics[width=.10\textwidth]{ci.png}};
\node[inner sep=0pt] (cs1) at (8.0,11) {\includegraphics[width=.10\textwidth]{cs.png}};
\node[inner sep=0pt] (as1) at (13.0,11) {\includegraphics[width=.10\textwidth]{as.png}};
\scalebox{0.95}{\scriptsize 
\draw [dashed] (5.9,1.8) -- (5.9,11.0);
\draw [dashed] (12,1.8) -- (12,11.0);
\node [right=-0.8cm] at (25pt,280pt){\textbf{Login protocol:}};
\node [right=-0.8cm] at (25pt,268pt){Inputs $UID_i$, $MID_i$ and $PW_i$};
\node [right=-0.8cm] at (25pt,256pt){Replaces $UID_i$ with $UP_{C_i}^{CS}$ and};
\node [right=0.4cm] at (25pt,242.5pt){$MID_i$ with $MP_{C_i}^{CS}$};
\node [right=-0.8cm] at (25pt,232pt) {Generates $TS_{C_i}$};
\node [right=-0.8cm] at (25pt,220pt) {Calls $GM$ and $CM$};
\node [right=-0.8cm] at (25pt,208pt) {$C_iK_{pr_i}=tmpC_iK_{pr_i}\oplus GM\oplus PW_i\oplus N_{C_i}$};
\node [right=-0.8cm] at (25pt,196pt) {Generates $N_{C_i}$};
\node [right=-0.8cm] at (25pt,184pt) {$C_iSig_1 = h(CM||N_{C_i}||TS_{C_i})$};
\node [right=-0.8cm] at (25pt,172pt) {$C_iSig_2= h(GM||N_{C_i}||TS_{C_i}||C_iSig_1||$};
\node [right=0.7cm] at (25pt,160pt) {$UP_{C_i}^{CS}||MP_{C_i}^{CS}||PW_i)$};
\node [right=-0.8cm] at (25pt,147pt) {Computes $tmpPW_i =PW_i\oplus N_{C_i}\oplus$};
\node [right=2.0cm] at (25pt,136pt) {$GM\oplus C_iSig_1$};
\node [right=-0.8cm] at (25pt,123pt) {$Request_i=Enc_i (TS_{C_i}||N_{C_i}|| C_iSig_1||UP_{C_i}^{CS}$};
\node [right=0.7cm] at (25pt,111pt) {$||MP_{C_i}^{CS}||tmpPW_i||C_iSig_2)$};
\node [right=-0.8cm] at (25pt,99pt) {$tmpC_iK_{pr_i}= C_iK_{pr_i}\oplus PW_i\oplus C_iSig_2$};

\draw [->,>=stealth] (.5,2.8) -- (5.5,2.8) node[above,pos=0.20] {\textbf{Login request}};
\node [right=5cm] at (25pt,185pt){\textbf{From $C_i$: (login request)}};
\node [right=5cm] at (25pt,173pt){$Dec_i (Request_i)$};
\node [right=5cm] at (25pt,161pt){Checks $TS_{CS}-T_{C_i}\leq \triangle T$};
\node [right=5cm] at (25pt,149pt){Checks $UP_{C_i}^{CS}$ and $MP_{C_i}^{CS}$ in datasets};
\node [right=5cm] at (25pt,137pt){Retrieves $GM$ from dataset};
\node [right=5cm] at (25pt,125pt){$CSSig_1 = h("Real\_MAC"||N_{C_i}||TS_{C_i})$};
\node [right=5cm] at (25pt,113pt){Checks $CSSig_1= C_iSig_1$};
\node [right=5cm] at (25pt,101pt){Computes $PW_i =tmpPW_i\oplus N_{C_i}\oplus GM$};
\node [right=7.2cm] at (25pt,89pt){$\oplus CSSig_1$};
\node [right=5cm] at (25pt,77pt){$CSSig_2 = h(GM||N_{C_i}||TS_{C_i}||CSSig_1||$};
\node [right=6.6cm] at (25pt,65pt){$UP_{C_i}^{CS}||MP_{C_i}^{CS}||PW_i)$};
\node [right=5cm] at (25pt,53pt){Checks $CSSig_2= C_iSig_2$};

} 
\end{tikzpicture}
	\caption{Login protocol}
	\label{fig:login_protocol}
\end{figure}

Patients and HC providers ($C_i$) should complete the registration and login protocol with $CS$ to become legitimate users of HC applications. Without this protocol, the user cannot complete the authentication process in RAMHU. User registration is performed once, namely, the user does not need to complete the registration protocol the next times (only the login protocol), the registration information has been kept in the servers until the revocation protocol and deletion the user security parameters such as pseudonyms, MAC address and $PW_i$, this protocol accomplishes the following steps (Figure~\ref{fig:registration_and_login_protocol} shows registration, and login protocol and Figure~\ref{fig:login_protocol} shows login protocol):\\

\noindent$C_{i}$ side:
\begin{itemize}
\item The user enters $UID_{i}$ (such as his/her name), $MID_i$ (such as medical centre name), $PW_{i}$ and $OTP_{i}$ for registration and login while only entering $UID_{i}$, $MID_i$, and $PW_{i}$ for the login protocol the next times to the client ($C_{i}$) application. $C_{i}$ replaces $UID_{i}$ with user's pseudonym ($UP^{CS}_{C_{i}}$), and $MID_i$ with medical centre pseudonym ($MP^{CS}_{C_{i}}$) to protect the authentication information when moving from $C_{i}$ to $CS$. $C_{i}$ generates timestamp ($TS_{C_{i}}$) to be used to verify the sending time of the registration/login request in $CS$. 
\item $C_{i}$ gets a MAC address (GM) by entering its IP (Internet protocol) address. The process of checking MAC ($CM$) is performed by $C_{i}$ to test the credibility of the MAC address. In the Linux system, we used the command "ethtool -P interface name" (such as "ethtool -P wlo1") in the $C_{i}$ application. If the result is identical to $GM$, it means that the MAC address is native ($CM$ = "$Real\_MAC$"). In Windows system, $C_{i}$ searches for string value "NetworkAddress" in the path of the system registry ("$HKEY\_LOCAL\_MACHINE\setminus SYSTEM\setminus CurrentControlSet\setminus Control\setminus Class\setminus \{4D36E972-E325-11CE-BFC1-$\ $08002BE10318\}\setminus$"). If NetworkAddress = null, $CM$ = "$Real\_MAC$"; otherwise $CM$ = "$Fake\_MAC$". The $GM$ send is encrypted with an registration/login request while $CM$ is implicitly sent with the signature value ($C_iSig_1$). If only login protocol, $C_i$ needs to extract private key from temporary key, MAC address, password, and random nonce through $tmpC_iK_{pr_i}\oplus GM\oplus PW_i\oplus N_{C_i}$. $C_{i}$ generates a random nonce ($N_{C_i}$) to change signature and encryption data and add anonymity to the registration/login request.
\item $C_{i}$ performs two signatures using the PHOTON-256 algorithm to protect information from modification. The first signature includes the parameters check MAC, nonce, and timestamp ($CiSig_{1} = h (CM\|N_{C_{i}}\|TS_{C_{i}}$). The second signature includes all the authentication parameters of the get MAC, nonce, timestamp, first signature, pseudonyms, one time password, and password ($CiSig_{2} = h(GM\|N_{C_{i}}\|TS_{C_{i}}\|$\ $CiSig_{1}\|UP^{CS}_{C_{i}}\|MP^{CS}_{C_{i}}$\ $\|OTP\|PW_{i}$). In the login protocol, $OTP$ is not added to the signature.
\item $C_i $ computes a temporary value ($tmpPW_i = PW_i\oplus N_ {C_i}\oplus GM\oplus C_iSig_1$) of $PW_i$ when moving from $C_i$ to $CS$.
\item $C_{i}$ uses ECIES to encrypt and hide all the data of this request ($Enc_{i}(TS_{C_{i}}\|N_{C_{i}}\|GM\|CiSig_{1}\|$\ $UP^{CS}_{C_{i}}\|MP^{CS}_{C_{i}}\|OTP_i\|tmpPW_{i}\|CiSig_{2})$). In the login protocol, $OTP_i$ and $GM$ are not added to the encryption as in Figure~\ref{fig:login_protocol}. After that, $C_{i}$ sends the registration and login request or login to $CS$ to complete the authentication protocol. Then $C_i$ hides the private key by $tmpC_iK_{pr_i}= C_iK_{pr_i}\oplus PW_i\oplus C_iSig_2$ . 
\end{itemize}

\noindent$CS$ side:
\begin{itemize}
\item Upon receiving a registration and login request or login request, $CS$ decrypts this request using ECIES's $C_{i}K_{pu_i}$ and $CSK_{pr_i}$. It checks timestamp ($TS_{CS}$-$TS_{C_{i}}$ $\leq$ $\triangle T$) to make sure that this request arrived at an appropriate time and without delay.
\item In the registration and login protocol, $CS$ examines the random $OTP_{i}$ in the dataset. If $OTP_{i}$ exists, then the user is legitimate for the registration process. After that, $CS$ deletes $OTP_{i}$ from the dataset to prevent it from being used the next times. If $OTP_{i}$ is not found, It discards the connection. If the login protocol, $CS$ examines the $UP_{C_i}^{CS}$ and $MP_{C_i}^{CS}$ and then tests their association with the MAC address in the dataset. If $GM$ is found, $CS$ completes the steps of this protocol; otherwise, it cancels the connection.
\item $CS$ needs to ensure that the user's device is legitimate within the network. $CS$ computes the signature value to make sure that the MAC address is native and non-modified ($CSSig_{1} = h ("Real\_MAC"\|N_{C_{i}}$\ $\|TS_{C_{i}})$. It examines the result of the computed signature ($CSSig_{1}$) with the  $C_{i}$'s signature ($C_{i}Sig_{1}$). If the signatures result are identical, then the device is legitimate and the MAC address did not change. In the registration and login protocol, $CS$ stores this address in dataset for use and check the next times in the login protocol.
\item $CS$ performs the computation operation the $tmpPW_i\oplus N_{C_i}\oplus GM\oplus CSSig_1$ to extract the $PW_i$ value and then uses this value to produce a second signature ($CSSig_2$).
\item $CS$ computes a second signature operation ($CSSig2$=\ $h (GM\|N_{C_{i}}\|TS_{C_{i}}\|CSSig_{1}$\ $\|UP_{C_{i}}^{CS}$\ $\|MP_{C_{i}}^{CS}\|OTP_{i}\|PW_{i}$) to guarantee that all the encrypted information is not changed. Then, it compares the computed signature ($CSSig_{2}$) with the received signature in the request ($CiSig_{2}$). If the signatures are identical, then the information for this request is unchanged or not tampered. In the login protocol, $OTP_i$ and $GM$ are not added to the signature. At this point, $CS$ prepares to send the user's authentication request to $AS$.
\end{itemize}

\subsubsection{Authentication Protocol} 
\begin{figure}[!t]
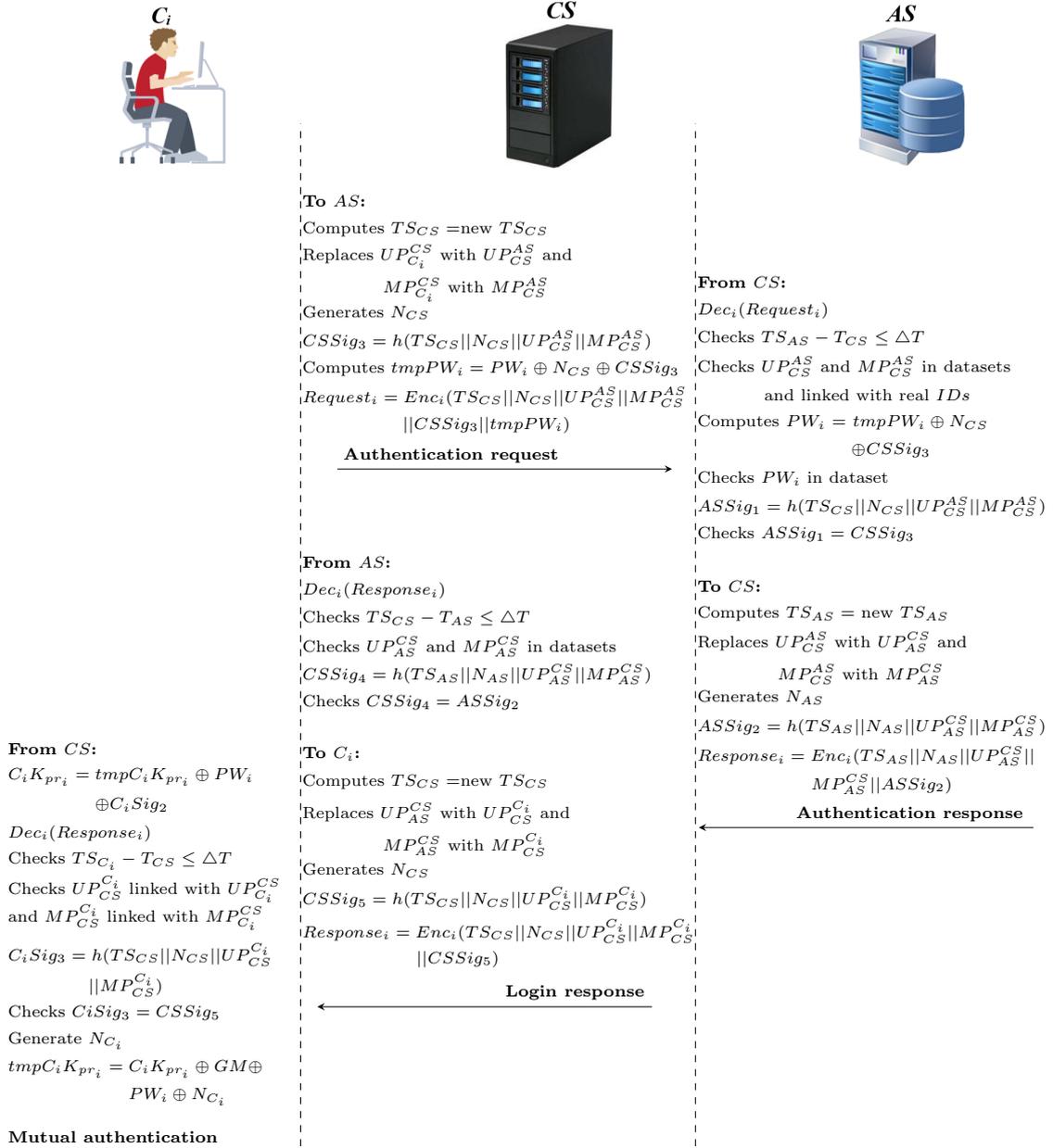

\centering
\scriptsize 
\begin{tikzpicture}
\node[inner sep=0pt] (u1) at (2.5,11) {\includegraphics[width=.10\textwidth]{ci.png}};
\node[inner sep=0pt] (cs1) at (8.0,11) {\includegraphics[width=.10\textwidth]{cs.png}};
\node[inner sep=0pt] (as1) at (13.0,11) {\includegraphics[width=.10\textwidth]{as.png}};
\scalebox{0.95}{\scriptsize 
\draw [dashed] (4.55,-4.4) -- (4.55,11);
\draw [dashed] (10.45,-4.4) -- (10.45,11);
\node [right=3.6cm] at (25pt,280pt){\textbf{To $AS$:}};
\node [right=3.6cm] at (25pt,268pt){Computes $TS_{CS}=$new $TS_{CS}$};
\node [right=3.6cm] at (25pt,256pt){Replaces $UP_{C_i}^{CS}$ with $UP_{CS}^{AS}$ and};
\node [right=4.8cm] at (25pt,242pt){$MP_{C_i}^{CS}$ with $MP_{CS}^{AS}$};
\node [right=3.6cm] at (25pt,232pt){Generates $N_{CS}$};
\node [right=3.6cm] at (25pt,220pt){$CSSig_3 = h(TS_{CS}||N_{CS}||UP_{CS}^{AS}||MP_{CS}^{AS})$};
\node [right=3.6cm] at (25pt,208pt){Computes $tmpPW_i = PW_i\oplus N_{CS}\oplus CSSig_3$};
\node [right=3.6cm] at (25pt,196pt){$Request_i=Enc_i(TS_{CS}||N_{CS}||UP_{CS}^{AS}||MP_{CS}^{AS}$};
\node [right=5.1cm] at (25pt,184pt){$||CSSig_3||tmpPW_i)$};

\draw [->,>=stealth] (5.1,5.8) -- (10.1,5.8) node[above,pos=0.34] {\textbf{Authentication request}};

\node [right=9.5cm] at (25pt,245pt){\textbf{From $CS$:}};
\node [right=9.5cm] at (25pt,233pt){$Dec_i (Request_i)$};
\node [right=9.5cm] at (25pt,221pt){Checks $TS_{AS}-T_{CS}\leq \triangle T$};
\node [right=9.5cm] at (25pt,209pt){Checks $UP_{CS}^{AS}$ and $MP_{CS}^{AS}$ in datasets};
\node [right=10.5cm] at (25pt,197pt){and linked with real $IDs$};
\node [right=9.5cm] at (25pt,185pt){Computes $PW_i$ = $tmpPW_i\oplus  N_{CS}$};
\node [right=11.8cm] at (25pt,173pt){$\oplus CSSig_3$};
\node [right=9.5cm] at (25pt,161pt){Checks $PW_i$  in dataset};
\node [right=9.5cm] at (25pt,149pt){$ASSig_1 = h(TS_{CS}||N_{CS}||UP_{CS}^{AS}||MP_{CS}^{AS})$};
\node [right=9.5cm] at (25pt,137pt){Checks $ASSig_1= CSSig_3$};

\node [right=9.5cm] at (25pt,115pt){\textbf{To $CS$:}};
\node [right=9.5cm] at (25pt,103pt){Computes $TS_{AS}=$ new $TS_{AS}$};
\node [right=9.5cm] at (25pt,91pt){Replaces $UP_{CS}^{AS}$ with $UP_{AS}^{CS}$ and};
\node [right=10.7cm] at (25pt,77pt){$MP_{CS}^{AS}$ with $MP_{AS}^{CS}$};
\node [right=9.5cm] at (25pt,67pt){Generates $N_{AS}$};
\node [right=9.5cm] at (25pt,55pt){$ASSig_2= h(TS_{AS}||N_{AS}||UP_{AS}^{CS}||MP_{AS}^{CS})$};
\node [right=9.5cm] at (25pt,42pt){$Response_i=Enc_i(TS_{AS}||N_{AS}||UP_{AS}^{CS}||$};
\node [right=11.2cm] at (25pt,30pt){$MP_{AS}^{CS}||ASSig_2)$};

\draw [->,>=stealth] (15.5,0.4) -- (10.5,0.4) node[above,pos=0.37] {\textbf{Authentication response}};

\node [right=3.6cm] at (25pt,125pt){\textbf{From $AS$:}};
\node [right=3.6cm] at (25pt,113pt){$Dec_i (Response_i)$};
\node [right=3.6cm] at (25pt,101pt){Checks $TS_{CS}-T_{AS}\leq \triangle T$};
\node [right=3.6cm] at (25pt,89pt){Checks $UP_{AS}^{CS}$ and $MP_{AS}^{CS}$ in datasets};
\node [right=3.6cm] at (25pt,77pt){$CSSig_4 = h(TS_{AS}||N_{AS}||UP_{AS}^{CS}||MP_{AS}^{CS})$};
\node [right=3.6cm] at (25pt,65pt){Checks $CSSig_4= ASSig_2$};

\node [right=3.6cm] at (25pt,43pt){\textbf{To $C_i$:}};
\node [right=3.6cm] at (25pt,31pt){Computes $TS_{CS}=$new $TS_{CS}$};
\node [right=3.6cm] at (25pt,18pt){Replaces $UP_{AS}^{CS}$ with $UP_{CS}^{C_i}$ and};
\node [right=4.8cm] at (25pt,4pt){$MP_{AS}^{CS}$ with $MP_{CS}^{C_i}$};
\node [right=3.6cm] at (25pt,-7pt){Generates $N_{CS}$};
\node [right=3.6cm] at (25pt,-19pt){$CSSig_5 = h(TS_{CS}||N_{CS}||UP_{CS}^{C_i}||MP_{CS}^{C_i})$};
\node [right=3.6cm] at (25pt,-33pt){$Response_i=Enc_i(TS_{CS}||N_{CS}||UP_{CS}^{C_i}||MP_{CS}^{C_i}$};
\node [right=5.3cm] at (25pt,-45pt){$||CSSig_5)$};
\draw [->,>=stealth] (9.8,-2.3) -- (4.8,-2.3) node[above,pos=0.23] {\textbf{Login response}};

\node [right=-0.8cm] at (25pt,45pt){\textbf{From $CS$:}};
\node [right=-0.8cm] at (25pt,33pt) {$C_iK_{pr_i}= tmpC_iK_{pr_i}\oplus PW_i$};
\node [right=0.5cm] at (25pt,21pt) {$\oplus C_iSig_2$};
\node [right=-0.8cm] at (25pt,9pt){$Dec_i (Response_i)$};
\node [right=-0.8cm] at (25pt,-03pt){Checks $TS_{C_i}-T_{CS}\leq \triangle T$};
\node [right=-0.8cm] at (25pt,-15pt){Checks $UP_{CS}^{C_i}$ linked with $UP_{C_i}^{CS}$};
\node [right=-0.8cm] at (25pt,-27pt){and $MP_{CS}^{C_i}$ linked with $MP_{C_i}^{CS}$};
\node [right=-0.8cm] at (25pt,-43pt) {$C_iSig_3 = h(TS_{CS}||N_{CS}||UP_{CS}^{C_i}$};
\node [right=0.4cm] at (25pt,-56pt) {$||MP_{CS}^{C_i })$};
\node [right=-0.8cm] at (25pt,-68pt) {Checks $CiSig_3= CSSig_5$};
\node [right=-0.8cm] at (25pt,-80pt) {Generate $N_{C_i}$};
\node [right=-0.8cm] at (25pt,-92pt) {$tmpC_iK_{pr_i}= C_iK_{pr_i}\oplus GM\oplus$};
\node [right=1.0cm] at (25pt,-104pt) {$PW_i\oplus N_{C_i}$};

\node [right=-0.8cm] at (25pt,-121pt) {\textbf{Mutual authentication}};

} 
\end{tikzpicture}
	\caption{Authentication protocol}
	\label{fig:authentication_protocol}
\end{figure}

The authentication protocol verifies the reliability of the users' security parameters with their personal information in the server. In this protocol, RAMHU needs to link pseudonyms and passwords with real information for users in $AS$'s datasets. Note that,  the users' information (such as name, age, address, mobile number, and passwords) are stored in a separate server (AS) and multi pseudonyms used to prevent detection and tracking of users' information. This protocol is illustrated in the following steps (Figure~\ref{fig:authentication_protocol} shows authentication protocol in RAMHU):\\

\noindent$CS$ side:
\begin{itemize}
\item $CS$ computes a new timestamp ($TS_{CS}$) to ensure the fresh authentication request.
\item $CS$ replaces $C_{i}$'s pseudonyms ($UP_{C_{i}}^{CS}$ and $MP_{C_{i}}^{CS}$) with $CS$'s pseudonyms ($UP_{CS}^{AS}$ and $MP_{CS}^{AS}$) to prevent attackers from tracking the authentication request. It generates random nonce ($N_{CS}$) to ensure an anonymous authentication request.
\item $CS$ signs security parameters by PHOTON-256 ($CSSig_{3} = h (TS_{CS}\|N_{CS}\|UP_{CS}^{AS}\|MP_{CS}^{AS}$) to prevent modification of authentication request information.
\item $CS$ computes temporary $PW_i$ by the computation of $PW_i\oplus N_{CS}\oplus CSSig3$.
\item $CS$ encrypts information of authentication request ($Enc_i(TS_{CS}\|N_{CS}\|UP_{CS}^{AS}\|MP_{CS}^{AS}\|CSSig_3$\ $\| tmpPW_i)$). It sends an authentication request to verify user's information in $AS$.
\end{itemize}

\noindent$AS$ side:
\begin{itemize}
\item Upon receiving the authentication request, $AS$ decrypts that request using ECIES's $ASK_{pr_{i}}$ and $CSK_{pu_{i}}$ to obtain the authentication information clearly.
\item It checks timestamp by ($TS_{AS}- TS_{CS} \leq \triangle T$) to ensure that the authentication request is not delayed. $AS$ checks the pseudonyms ($UP_{CS}^{AS}$ and $MP_{CS}^{AS}$) sent from CS and correlates it with the user's real identifier ($UID_i$ and $MID_i$) in the datasets. $AS$ extracts user's password from the equation $PW_i= tmpPW_i\oplus N_{CS}\oplus CSSig_3$. After that, It checks matching $PW_i$ in the dataset. $AS$ computes the value of the signature based on authentication information ($ASSig_1 = h (TS_{CS}\|N_{CS}\|UP_{CS}^{AS}\|MP_{CS}^{AS}$)) by PHOTON-256. $AS$ compares the computed value of the signature ($ASSig_1$) with the value of the received signature ($CSSig_3$). If the signature values are identical, the user's information in the request for the signature is unmodified. At this point, $AS$ considers this user to be legitimate and reliable. 
\item $AS$ prepares a response to authenticate the request of that user. It computes new timestamp ($TS_{AS}$ = new $TS_{AS}$) to prevent delayed or replayed requests at later times. $AS$ replaces $CS$'s pseudonyms ($UP_{CS}^{AS}$ and $UP_{CS}^{AS}$) received with $AS$'s pseudonyms ($UP_{AS}^{CS}$ and $MP_{AS}^{CS}$) to hide users' information. It generates a new random nonce ($N_{AS}$) to add anonymity and prevent attacks from encryption and signature analysis.
\item $AS$ computes a signature ($ASSig_2 = h (TS_{AS}\|N_{AS}\|UP_{AS}^{CS}\|MP_{AS}^{CS})$) to prevent modifications of authentication response information.
\item $AS$ encrypts the authentication information ($Enc_i(TS_{AS}\|N_{AS}\|UP_{AS}^{CS}\|MP_{AS}^{CS}\|$\ $ASSig_2)$) and sends the authentication response to $CS$ to complete the authentication process.
\end{itemize}

\noindent$CS$ side:
\begin{itemize}
\item $CS$ decrypts the authentication response ($Enc_i$) received from $AS$. It checks the timestamp value ($TS_{CS}- TS_{AS}  \leq \triangle T$) to prevent late authentication responses. It examines that $UP_{CS}^{AS}$ and $MP_{CS}^{AS}$ in datasets to complete the process of linking multi pseudonyms to the user.
\item $CS$ computes the signature $CSSig4 = h (TS_{AS}\|N_{AS}\|UP_{AS}^{CS}\|MP_{AS}^{CS})$ for authentication response information. It compares the computed result of the signature ($CSSig_4$) with the value of the received signature ($ASSig_2$). If the signature values match, then the authentication response information is unchanged.
\item After this point, $CS$ initiates a login response request. It computes the value of new timestamp ($TS_{CS} = new TS_{CS}$). It replaces $AS$'s pseudonyms ($UP_{AS}^{CS}$ and $MP_{AS}^{CS}$) received with $UP_{CS}^{C_i}$ and $MP_{CS}^{C_i}$. It generates a new random nonce ($N_{CS}$) to hide encryption and signature information.
\item $CS$ computes a new signature value ($CSSig_5 = h (TS_{CS}\|N_{CS}\|UP_{CS}^{C_i}\|MP_{CS}^{C_i})$) to protect login response information from modification.
\item $CS$ encrypts login response information ($Enc_i(TS_{CS}\|N_{CS}\|UP_{CS}^{C_i}\|MP_{CS}^{C_i}\|CSSig_5)$) and sends this response to $C_i$.
\end{itemize}

\noindent$C_i$ side:
\begin{itemize}
\item $C_i$ extracts his private key ($C_iK_{pr_i}$) from the $tmpC_iK_{pr_i}\oplus PW_i \oplus C_iSig_2$, and decrypts the login request ($Enc_i$) received from $CS$.
\item $C_i$ checks timestamp ($TS_{C_i} - TS_{CS} \leq \triangle T$) to ensure that the login response is not late or replayed.
\item  $C_i$ checks that $CS$'s pseudonyms ($UP_{CS}^{C_i}$ and $MP_{CS}^{C_i}$) are received in the dataset and associated with $UP_{C_i}^{CS}$ and $MP_{C_i}^{CS}$. At this point, RAMHU applied multi pseudonyms among model entities ($Ci$, $CS$, and $AS$) to prevent traceability in linking the real information of the user with pseudonyms.
\item $C_i$ calculates the value of the signature $CiSig_3 = h (TS_{CS}\|N_{CS}\|UP_{CS}^{C_i}\|MP_{CS}^{C_i})$ for login response information. It compares the result of the computed signature ($CiSig_3$) and the value of the received signature ($CSSig_5$). If the signature values are identical, namely, the login response information is unmodified or not tampered, $C_i$ accepts the login response; otherwise $C_i$ discards the login response. Then, $C_i$ hides its private key by $C_iK_{pr_i}\oplus GM\oplus PW_i\oplus N_{C_i}$ after generating a random nonce to prevent the detection of the private key if the device is hacked. At this point, if all processes are achieved correctly, then all requests are considered legitimate and reliable through the implementation of mutual authentication.
\end{itemize}

\subsubsection{Password Update Protocol} 
\begin{figure}[!t]
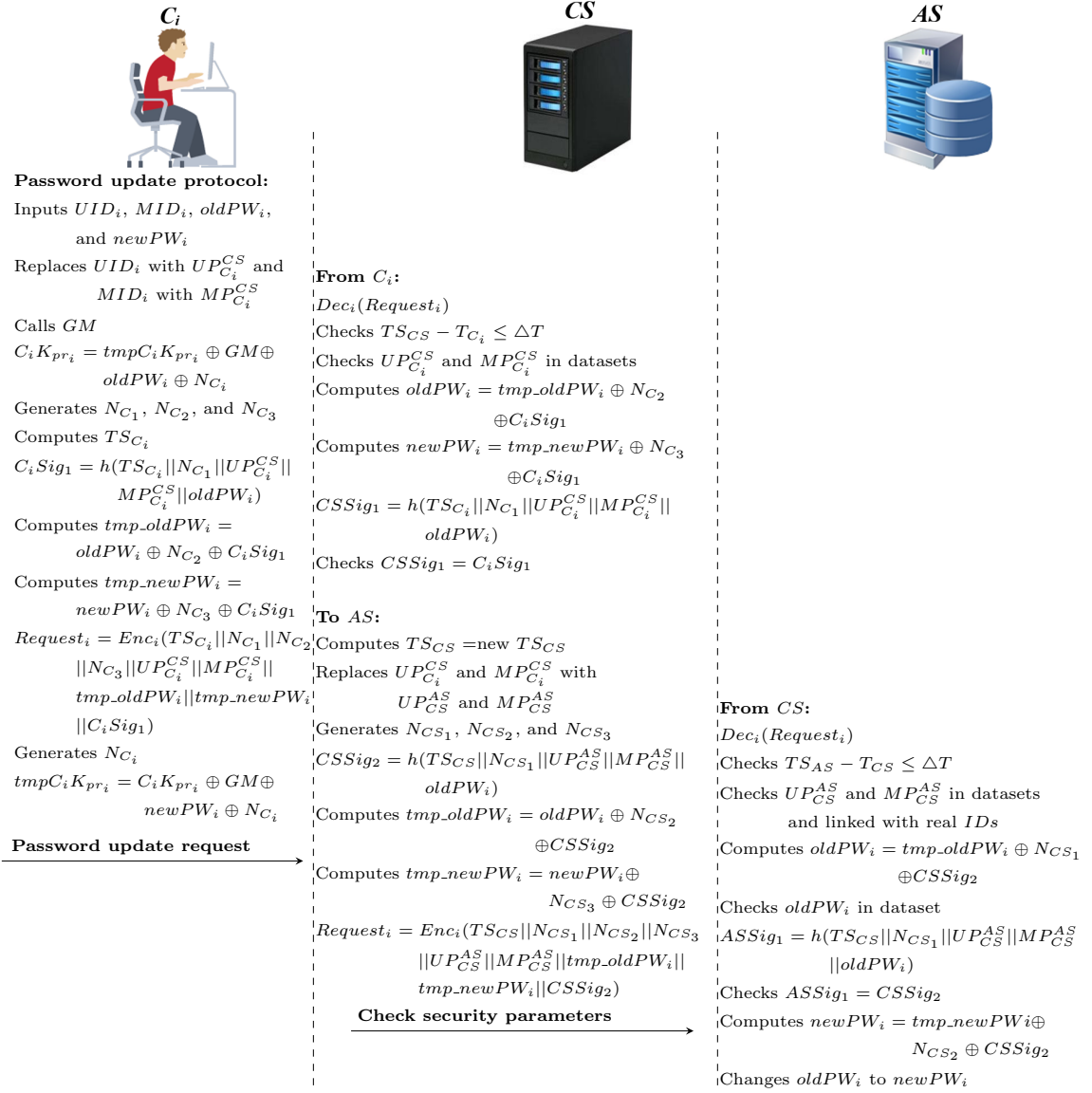

\centering
\scriptsize 
\begin{tikzpicture}
\node[inner sep=0pt] (u1) at (2.5,11) {\includegraphics[width=.10\textwidth]{ci.png}};
\node[inner sep=0pt] (cs1) at (8.0,11) {\includegraphics[width=.10\textwidth]{cs.png}};
\node[inner sep=0pt] (as1) at (13.0,11) {\includegraphics[width=.10\textwidth]{as.png}};
\scalebox{0.95}{\scriptsize 
\draw [dashed] (4.55,-3.1) -- (4.55,11);
\draw [dashed] (10.45,-3.1) -- (10.45,11);
\node [right=-0.8cm] at (25pt,290pt){\textbf{Password update protocol:}};
\node [right=-0.8cm] at (25pt,278pt){Inputs $UID_i$, $MID_i$, $oldPW_i$,};
\node [right=0.1cm] at (25pt,266pt){and $newPW_i$};
\node [right=-0.8cm] at (25pt,254pt){Replaces $UID_i$ with $UP_{C_i}^{CS}$ and};
\node [right=0.4cm] at (25pt,242pt){$MID_i$ with $MP_{C_i}^{CS}$};
\node [right=-0.8cm] at (25pt,230pt){Calls $GM$ };
\node [right=-0.8cm] at (25pt,218pt){$C_iK_{pr_i}= tmpC_iK_{pr_i}\oplus GM\oplus$ };
\node [right=0.5cm] at (25pt,206pt){$oldPW_i\oplus N_{C_i}$ };
\node [right=-0.8cm] at (25pt,194pt) {Generates $N_{C_1}$, $N_{C_2}$, and $N_{C_3}$};
\node [right=-0.8cm] at (25pt,182pt) {Computes $TS_{C_i}$};
\node [right=-0.8cm] at (25pt,170pt) {$C_iSig_1 = h(TS_{C_i}||N_{C_1}||UP_{C_i}^{CS}||$};
\node [right=0.7cm] at (25pt,158pt) {$MP_{C_i}^{CS}||oldPW_i)$};
\node [right=-0.8cm] at (25pt,146pt) {Computes $tmp\_oldPW_i=$};
\node [right=0.1cm] at (25pt,134pt) {$oldPW_i\oplus N_{C_2} \oplus C_iSig_1$};
\node [right=-0.8cm] at (25pt,122pt) {Computes $tmp\_newPW_i=$};
\node [right=0.1cm] at (25pt,110pt) {$newPW_i\oplus N_{C_3}\oplus C_iSig_1$};
\node [right=-0.8cm] at (25pt,98pt) {$Request_i=Enc_i(TS_{C_i}||N_{C_1}||N_{C_2}$ };
\node [right=0.1cm] at (25pt,86pt) {$||N_{C_3}||UP_{C_i}^{CS}||MP_{C_i}^{CS}||$ };
\node [right=0.1cm] at (25pt,74pt) {$tmp\_oldPW_i||tmp\_newPW_i$};
\node [right=0.1cm] at (25pt,62pt) {$||C_iSig_1)$ };

\node [right=-0.8cm] at (25pt,50pt) {Generates $N_{C_i}$ };
\node [right=-0.8cm] at (25pt,38pt){$tmpC_iK_{pr_i}= C_iK_{pr_i}\oplus GM\oplus$ };
\node [right=1.1cm] at (25pt,26pt){$newPW_i\oplus N_{C_i}$ };

\draw [->,>=stealth] (0,0.2) -- (4.4,0.2) node[above,pos=0.43] {\textbf{Password update request}};

\node [right=3.6cm] at (25pt,250pt){\textbf{From $C_i$:}};
\node [right=3.6cm] at (25pt,238pt){$Dec_i (Request_i)$};
\node [right=3.6cm] at (25pt,226pt){Checks $TS_{CS}-T_{C_i}\leq \triangle T$};
\node [right=3.6cm] at (25pt,214pt){Checks $UP_{C_i}^{CS}$ and $MP_{C_i}^{CS}$ in datasets};
\node [right=3.6cm] at (25pt,202pt){Computes $oldPW_i= tmp\_oldPW_i\oplus N_{C_2}$};
\node [right=6.2cm] at (25pt,190pt){$\oplus C_iSig_1$};
\node [right=3.6cm] at (25pt,178pt){Computes $newPW_i= tmp\_newPW_i\oplus N_{C_3}$};
\node [right=6.4cm] at (25pt,166pt){$\oplus C_iSig_1$};
\node [right=3.6cm] at (25pt,154pt){$CSSig_1 = h(TS_{C_i}||N_{C_1}||UP_{C_i}^{CS}||MP_{C_i}^{CS}||$};
\node [right=5.2cm] at (25pt,142pt){$oldPW_i)$};
\node [right=3.6cm] at (25pt,130pt){Checks $CSSig_1= C_iSig_1$};

\node [right=3.6cm] at (25pt,108pt){\textbf{To $AS$:}};
\node [right=3.6cm] at (25pt,96pt){Computes $TS_{CS}=$new $TS_{CS}$};
\node [right=3.6cm] at (25pt,84pt){Replaces $UP_{C_i}^{CS}$ and $MP_{C_i}^{CS}$ with};
\node [right=4.8cm] at (25pt,72pt){$UP_{CS}^{AS}$ and $MP_{CS}^{AS}$};
\node [right=3.6cm] at (25pt,60pt){Generates $N_{CS_1}$, $N_{CS_2}$, and $N_{CS_3}$};
\node [right=3.6cm] at (25pt,48pt){$CSSig_2 = h(TS_{CS}||N_{CS_1}||UP_{CS}^{AS}||MP_{CS}^{AS}||$};
\node [right=5.2cm] at (25pt,36pt){$oldPW_i)$};
\node [right=3.6cm] at (25pt,24pt){Computes $tmp\_oldPW_i= oldPW_i\oplus N_{CS_2}$};
\node [right=6.8cm] at (25pt,12pt){$\oplus CSSig_2$};
\node [right=3.6cm] at (25pt,0pt){Computes $tmp\_newPW_i= newPW_i\oplus$};
\node [right=7.0cm] at (25pt,-12pt){$N_{CS_3}\oplus CSSig_2$};
\node [right=3.6cm] at (25pt,-24pt){$Request_i=Enc_i(TS_{CS}||N_{CS_1}||N_{CS_2}||N_{CS_3}$};
\node [right=5.1cm] at (25pt,-36pt){$||UP_{CS}^{AS}||MP_{CS}^{AS}||tmp\_oldPW_i||$};
\node [right=5.1cm] at (25pt,-48pt){$tmp\_newPW_i||CSSig_2)$};
\draw [->,>=stealth] (5.1,-2.3) -- (10.1,-2.3) node[above,pos=0.39] {\textbf{Check security parameters}};

\node [right=9.5cm] at (25pt,70pt){\textbf{From $CS$:}};
\node [right=9.5cm] at (25pt,58pt){$Dec_i (Request_i)$};
\node [right=9.5cm] at (25pt,46pt){Checks $TS_{AS}-T_{CS}\leq \triangle T$};
\node [right=9.5cm] at (25pt,34pt){Checks $UP_{CS}^{AS}$ and $MP_{CS}^{AS}$ in datasets};
\node [right=10.5cm] at (25pt,22pt){and linked with real $IDs$};
\node [right=9.5cm] at (25pt,10pt){Computes $oldPW_i = tmp\_oldPW_i\oplus N_{CS_1}$};
\node [right=12.1cm] at (25pt,-1pt){$\oplus CSSig_2$};
\node [right=9.5cm] at (25pt,-14pt){Checks $oldPW_i$  in dataset};

\node [right=9.5cm] at (25pt,-26pt){$ASSig_1 = h(TS_{CS}||N_{CS_1}||UP_{CS}^{AS}||MP_{CS}^{AS}$};
\node [right=11.1cm] at (25pt,-38pt){$||oldPW_i)$};
\node [right=9.5cm] at (25pt,-50pt){Checks $ASSig_1= CSSig_2$};
\node [right=9.5cm] at (25pt,-62pt){Computes $newPW_i= tmp\_newPWi\oplus $};
\node [right=12.3cm] at (25pt,-74pt){$N_{CS_2}\oplus CSSig_2$};
\node [right=9.5cm] at (25pt,-86pt){Changes $oldPW_i$ to $newPW_i$};

} 
\end{tikzpicture}
	\caption{Password update protocol}
	\label{fig:password_update_protocol}
\end{figure}

Updating the password is a security procedure to ensure  authentication of legitimate users. The process of changing $PW_i$ is important in any HC system for two reasons. First, preventing the use of $PW_i$ fixed for a long time and which reduces the guessing attacks. Second, changing $PW_i$ gives users more flexibility in choosing the appropriate $PW_i$. This process requires strict security measures to protect new $PW_i$. RAMHU provides the legitimate user with a mechanism to change his/her password at any time. If the user wants to change his/her $PW_i$, the following illustration describes the new $PW_i$ protect procedures in a secure manner (Figure~\ref{fig:password_update_protocol} shows password update protocol in RAMHU).

\begin{itemize}
\item $C_i$ side: $Ci$ enters $UID_i$, $MID_i$, old $PW_i$, and new $PW_i$. It replaces $UID_i$ and $MID_i$ with a pseudonyms to hide the user's real information. $C_i$ calls MAC address, then extracts the private key from the $tmpC_iK_{pr_i}\oplus GM\oplus oldPW_i\oplus N_{C_i}$ to use in the encryption process of the password update request. $C_i$ generates three random nonces ($N_{C_1}$, $N_{C_2}$, and $N_{C_3}$) and computes a new timestamp ($TS_{C_{i}}$). $C_i$ computes the signature value ($C_iSig_1 = h (TS_{C_{i}}\|N_{C_1}\|UP_{C_{i}}^{CS}\|MP_{C_{i}}^{CS}\|oldPW_i$) by PHOTON-256 based on the parameters of the password change request. It applies an anonymity mechanism to old $PW_i$ and new $PW_i$ ($tmp\_oldPW_i$ = $oldPW_i\oplus N_{C_2}\oplus CiSig_1$ and $tmp\_newPW_i$ = $newPW_i\oplus N_{C_3}\oplus CiSig_1$) to hide passwords and not explicitly send it in the $PW_i$ change request. It encrypts the $PW_i$ change request ($Enc_i(TS_{C_{i}}\|N_{C_1}\| N_{C_2}\| N_{C_3}\|\ UP_{C_{i}}^{CS}\|$\ $MP_{C_{i}}^{CS}\|$\ $tmp\_oldPW_i\|$\ $tmp\_newPW_i$\ $\|C_iSig_1)$) and sends it to $CS$, $C_i$ then hides its private key by $C_iK_{pr_i}\oplus GM\oplus newPW_i\oplus N_{C_i}$.

\item $CS$ side: $CS$ receives and decrypts a password update request ($Enc_i$). It examines timestamp to prevent delayed requests, and examines $UP_{C_{i}}^{CS}$ and $MP_{C_{i}}^{CS}$ in datasets. $CS$ extracts old $PW_i$ and new $PW_i$ from $tmp\_oldPW_i\oplus N_{C_2}\oplus CiSig_1$ and $tmp\_newPW_i$\ $\oplus N_{C_3}\oplus CiSig_1$. Then, it computes signature value ($CSSig_1$) depending on $TS_{C_i}, N_{C_1},$\ $UP_{C_{i}}^{CS},$\ $MP_{C_{i}}^{CS}$, and $oldPW_i$ to check the matching between $CSSig_1$ and $C_iSig_1$. Similarly, in authentication protocol, $CS$ computes $TS_{CS}$ and replaces pseudonyms. $CS$ generates three nonces $N_{CS_1}$, $N_{CS_2}$, and $N_{CS_3}$ and then $CS$ computes signature value ($h (TS_{C_{i}}\|N_{C_1}\|UP_{C_{i}}^{CS}\|MP_{C_{i}}^{CS}\|oldPW_i$) and hides old $PW_i$ and new $PW_i$ by $oldPW_i\oplus N_{CS_2}\oplus CSSig_2$ and $newPW_i\oplus N_{CS_3}\oplus CSSig_2$. It encrypts password update request with security parameters ($TS_{CS}$, $N_{CS_1}$, $N_{CS_2}$, $N_{CS_3}$, $UP_{CS}^{AS}$, $MP_{CS}^{AS}$, $tmp\_oldPW_i$, $tmp\_newPW_i$, and $CSSig2$) and sends to $AS$.

\item $AS$ side: $AS$ receives password update request, and decrypts ($Enc_i$) this request with $ASK_{pr_i}$, and $CSK_{pu_i}$. It checks time delay and then, it checks link pseudonyms with real information for users. It extracts old $PW_i$ from $tmp\_oldPW_i\oplus N_{CS_2}\oplus CSSig_2$ and then, it checks old $PW_i$ in dataset. It computes the signature value $ASSig_1 = h (TS_{CS}\|N_{CS_1}\|UP_{CS}^{AS}\|MP_{CS}^{AS}\|$\ $oldPW_i)$, and then, it compares the calculated result ($ASSig_1$) with the result of the received signature ($CSSig_2$). If identical, the signature is true; otherwise, $AS$ rejects the $PW_i$ change request. $AS$ performs the calculation $tmp\_newPW_i\oplus NCS_3\oplus CSSig_2$ to obtain the new $PW_i$ value. If all previous checks are validated, $AS$ changes old $PW_i$ to new $PW_i$.
\end{itemize}
\subsubsection{Revocation Protocol} 
\begin{figure}[!t]
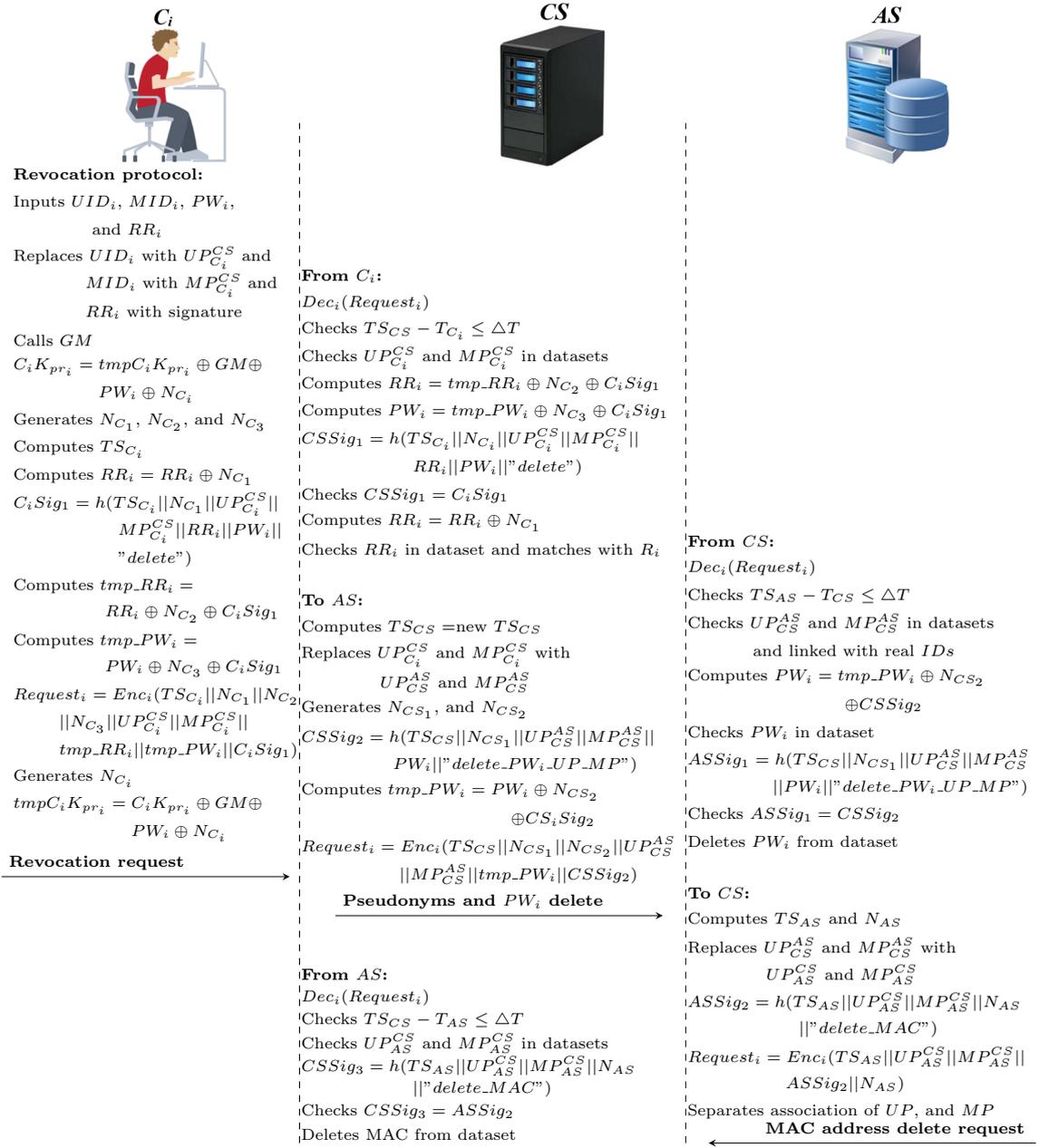

\centering
\scriptsize 
\begin{tikzpicture}
\node[inner sep=0pt] (u1) at (2.5,11) {\includegraphics[width=.10\textwidth]{ci.png}};
\node[inner sep=0pt] (cs1) at (8.0,11) {\includegraphics[width=.10\textwidth]{cs.png}};
\node[inner sep=0pt] (as1) at (13.0,11) {\includegraphics[width=.10\textwidth]{as.png}};
\scalebox{0.95}{\scriptsize 
\draw [dashed] (4.55,-4.7) -- (4.55,11);
\draw [dashed] (10.45,-4.7) -- (10.45,11);
\node [right=-0.8cm] at (25pt,290pt){\textbf{Revocation protocol:}};
\node [right=-0.8cm] at (25pt,278pt){Inputs $UID_i$, $MID_i$, $PW_i$,};
\node [right=0.4cm] at (25pt,266pt){and $RR_i$};
\node [right=-0.8cm] at (25pt,254pt){Replaces $UID_i$ with $UP_{C_i}^{CS}$ and};
\node [right=0.3cm] at (25pt,242pt){$MID_i$ with $MP_{C_i}^{CS}$ and};
\node [right=0.3cm] at (25pt,230pt){$RR_i$ with signature};
\node [right=-0.8cm] at (25pt,218pt){Calls $GM$ };
\node [right=-0.8cm] at (25pt,206pt){$C_iK_{pr_i}= tmpC_iK_{pr_i}\oplus GM\oplus$ };
\node [right=0.5cm] at (25pt,194pt){$PW_i\oplus N_{C_i}$ };
\node [right=-0.8cm] at (25pt,182pt) {Generates $N_{C_1}$, $N_{C_2}$, and $N_{C_3}$};
\node [right=-0.8cm] at (25pt,170pt) {Computes $TS_{C_i}$};
\node [right=-0.8cm] at (25pt,158pt) {Computes $RR_i = RR_i\oplus N_{C_1}$};
\node [right=-0.8cm] at (25pt,146pt) {$C_iSig_1 = h(TS_{C_i}||N_{C_1}||UP_{C_i}^{CS}||$};
\node [right=0.8cm] at (25pt,134pt) {$MP_{C_i}^{CS}||RR_i||PW_i||$};
\node [right=0.8cm] at (25pt,122pt) {$"delete")$};
\node [right=-0.8cm] at (25pt,110pt) {Computes $tmp\_RR_i=$};
\node [right=0.6cm] at (25pt,98pt) {$RR_i\oplus N_{C_2} \oplus C_iSig_1$};
\node [right=-0.8cm] at (25pt,86pt) {Computes $tmp\_PW_i=$};
\node [right=0.6cm] at (25pt,74pt) {$PW_i\oplus N_{C_3}\oplus C_iSig_1$};
\node [right=-0.8cm] at (25pt,62pt) {$Request_i=Enc_i(TS_{C_i}||N_{C_1}||N_{C_2}$ };
\node [right=-0.1cm] at (25pt,50pt) {$||N_{C_3}||UP_{C_i}^{CS}||MP_{C_i}^{CS}||$ };
\node [right=-0.1cm] at (25pt,38pt) {$tmp\_RR_i||tmp\_PW_i||C_iSig_1)$ };
\node [right=-0.8cm] at (25pt,26pt) {Generates $N_{C_i}$ };
\node [right=-0.8cm] at (25pt,14pt){$tmpC_iK_{pr_i}= C_iK_{pr_i}\oplus GM\oplus$ };
\node [right=1.0cm] at (25pt,2pt){$PW_i\oplus N_{C_i}$ };

\draw [->,>=stealth] (0,-0.6) -- (4.4,-0.6) node[above,pos=0.33] {\textbf{Revocation request}};

\node [right=3.6cm] at (25pt,246pt){\textbf{From $C_i$:}};
\node [right=3.6cm] at (25pt,234pt){$Dec_i (Request_i)$};
\node [right=3.6cm] at (25pt,222pt){Checks $TS_{CS}-T_{C_i}\leq \triangle T$};
\node [right=3.6cm] at (25pt,210pt){Checks $UP_{C_i}^{CS}$ and $MP_{C_i}^{CS}$ in datasets};
\node [right=3.6cm] at (25pt,198pt){Computes $RR_i= tmp\_RR_i\oplus N_{C_2}\oplus C_iSig_1$};            
\node [right=3.6cm] at (25pt,186pt){Computes $PW_i= tmp\_PW_i\oplus N_{C_3}\oplus C_iSig_1$};                       
\node [right=3.6cm] at (25pt,174pt){$CSSig_1 = h(TS_{C_i}||N_{C_i}||UP_{C_i}^{CS}||MP_{C_i}^{CS}||$};
\node [right=5.3cm] at (25pt,162pt){$RR_i||PW_i||"delete")$};
\node [right=3.6cm] at (25pt,150pt){Checks $CSSig_1= C_iSig_1$};
\node [right=3.6cm] at (25pt,138pt){Computes $RR_i = RR_i\oplus N_{C_1}$};
\node [right=3.6cm] at (25pt,126pt){Checks $RR_i$ in dataset and matches with $R_i$};

\node [right=3.6cm] at (25pt,104pt){\textbf{To $AS$:}};
\node [right=3.6cm] at (25pt,92pt){Computes $TS_{CS}=$new $TS_{CS}$};
\node [right=3.6cm] at (25pt,80pt){Replaces $UP_{C_i}^{CS}$ and $MP_{C_i}^{CS}$ with};
\node [right=4.8cm] at (25pt,68pt){$UP_{CS}^{AS}$ and $MP_{CS}^{AS}$};
\node [right=3.6cm] at (25pt,56pt){Generates $N_{CS_1}$, and $N_{CS_2}$};
\node [right=3.6cm] at (25pt,44pt){$CSSig_2 = h(TS_{CS}||N_{CS_1}||UP_{CS}^{AS}||MP_{CS}^{AS}||$};
\node [right=5cm] at (25pt,32pt){$PW_i||"delete\_PW_i\_UP\_MP")$};
\node [right=3.6cm] at (25pt,20pt){Computes $tmp\_PW_i= PW_i\oplus N_{CS_2}$};
\node [right=6.8cm] at (25pt,8pt){$\oplus CS_iSig_2$};
\node [right=3.6cm] at (25pt,-4pt){$Request_i=Enc_i(TS_{CS}||N_{CS_1}||N_{CS_2}||UP_{CS}^{AS}$};
\node [right=5.1cm] at (25pt,-16pt){$||MP_{CS}^{AS}||tmp\_PW_i||CSSig_2)$};

\draw [->,>=stealth] (5.1,-1.2) -- (10.1,-1.2) node[above,pos=0.42] {\textbf{Pseudonyms and $PW_i$ delete}};

\node [right=9.5cm] at (25pt,130pt){\textbf{From $CS$:}};
\node [right=9.5cm] at (25pt,118pt){$Dec_i (Request_i)$};
\node [right=9.5cm] at (25pt,106pt){Checks $TS_{AS}-T_{CS}\leq \triangle T$};
\node [right=9.5cm] at (25pt,94pt){Checks $UP_{CS}^{AS}$ and $MP_{CS}^{AS}$ in datasets};
\node [right=10.5cm] at (25pt,82pt){and linked with real $IDs$};
\node [right=9.5cm] at (25pt,70pt){Computes $PW_i = tmp\_PW_i\oplus N_{CS_2}$};
\node [right=11.9cm] at (25pt,58pt){$\oplus CSSig_2$};
\node [right=9.5cm] at (25pt,46pt){Checks $PW_i$  in dataset};
\node [right=9.5cm] at (25pt,34pt){$ASSig_1 = h(TS_{CS}||N_{CS_1}||UP_{CS}^{AS}||MP_{CS}^{AS}$};
\node [right=10.8cm] at (25pt,22pt){$||PW_i||"delete\_PW_i\_UP\_MP")$};
\node [right=9.5cm] at (25pt,10pt){Checks $ASSig_1= CSSig_2$};
\node [right=9.5cm] at (25pt,-2pt){Deletes $PW_i$ from dataset};

\node [right=9.5cm] at (25pt,-24pt){\textbf{To $CS$:}};
\node [right=9.5cm] at (25pt,-36pt){Computes $TS_{AS}$ and $N_{AS}$};
\node [right=9.5cm] at (25pt,-48pt){Replaces $UP_{CS}^{AS}$ and $MP_{CS}^{AS}$ with};
\node [right=10.7cm] at (25pt,-60pt){$UP_{AS}^{CS}$ and $MP_{AS}^{CS}$};
\node [right=9.5cm] at (25pt,-72pt){$ASSig_2= h (TS_{AS}||UP_{AS}^{CS}||MP_{AS}^{CS}||N_{AS}$};
\node [right=11.2cm] at (25pt,-84pt){$||"delete\_MAC")$};
\node [right=9.5cm] at (25pt,-96pt){$Request_i=Enc_i(TS_{AS}||UP_{AS}^{CS}||MP_{AS}^{CS}||$};
\node [right=11.0cm] at (25pt,-108pt){$ASSig_2||N_{AS})$};
\node [right=9.5cm] at (25pt,-120pt){Separates association of $UP$, and $MP$};

\draw [->,>=stealth] (15.8,-4.7) -- (10.8,-4.7) node[above,pos=0.43] {\textbf{MAC address delete request}};

\node [right=3.6cm] at (25pt,-60pt){\textbf{From $AS$:}};
\node [right=3.6cm] at (25pt,-70pt){$Dec_i (Request_i)$};
\node [right=3.6cm] at (25pt,-80pt){Checks $TS_{CS}-T_{AS}\leq \triangle T$};
\node [right=3.6cm] at (25pt,-90pt){Checks $UP_{AS}^{CS}$ and $MP_{AS}^{CS}$ in datasets};
\node [right=3.6cm] at (25pt,-100pt){$CSSig_3 = h(TS_{AS}||UP_{AS}^{CS}||MP_{AS}^{CS}||N_{AS}$};
\node [right=5.3cm] at (25pt,-110pt){$||"delete\_MAC")$};
\node [right=3.6cm] at (25pt,-120pt){Checks $CSSig_3= ASSig_2$};
\node [right=3.6cm] at (25pt,-130pt){Deletes MAC from dataset};

} 
\end{tikzpicture}
	\caption{Revocation protocol}
	\label{fig:revocation_protocol}
\end{figure}

Revocation in HC applications uses when a user finishes a task or to prevent attack. This protocol can be completed by $C_i$, or $AS$. If $C_i$ wants to cancel his account from the HC system after completing his/her duties such as a research doctor who uses the system for a limited period and then cancel his account after the completion of his duties. Additionally, $AS$ can revoke the account of any user who performs suspicious activities (internal attacks) that are not within his/her privileges such as a nurse who wants to access the personal information of a particular doctor, or patient. Furthermore, the user can request from authorities provider ($AS$) to revoke his/her account information that is associated with his/her data (note that the patients' data history remains stored in $DS$ even after the completion of the revocation protocol). The protocol of revocation is extremely important in restricting the malicious activities of HC systems. RAMHU includes a revocation protocol to provide strict security procedures in protecting users' authentication information (Figure~\ref{fig:revocation_protocol} shows revocation protocol in $C_i$ side in RAMHU):\\

\noindent Revocation from $C_i$ side:
\begin{itemize}
\item $C_i$ enters $UID_i$, $MID_i$, and $PW_i$ and then replaces $UID_i$ with $UP_{C_i}^{CS}$ and $MID_i$ with $MP_{C_i}^{CS}$. It chooses the revocation reason ($RR_i$) from the drop-down list (such as ending the researcher's study, ending a satisfactory condition, resigning a professional, changing a health institution, and the unwillingness of a patient to use the system). These reasons have converted to signatures using PHOTON to get  MDs with a 256-bits in the dataset. $C_i$ computes new $TS_{C_i}$, $N_{C_1}$, $N_{C_2}$, and $N_{C_3}$. Then, $C_i$ performs the process of $RR_i \oplus N_{C_1}$ to add randomness for $RR_i$. Using this procedure is very useful in tightening security and distinguishing the roles of users (patients or professionals) in $CS$. It computes the signature $C_iSig_1 = h (TS_{C_i}\|N_{C_1}\|UP_{C_i}^{CS}\|MP_{C_i}^{CS}\|$\ $RR_i\|$\ $PW_i\|$\ $"delete")$. $C_i$ performs a computation to hide the $RR_i$ ($tmpRR_i = RR_i\oplus N_{C_2}\oplus C_iSig_1$). $C_i$ computes the temporary $PW_i$ value of $PW_i\oplus N_ {C_3}\oplus C_iSig_1$ to hide the $PW_i$ value. Additionally, It computes encryption ($Enc_i(TS_{C_i}\|N_{C_1}\|N_{C_2}\|N_{C_3}\|UP_{C_i}^{CS}\|MP_{C_i}^{CS}\|$\ $tmpRR_i\|$\ $tmpPW_i$\ $\|C_iSig_1)$) and sends a revocation request to $CS$. Then, It hides a private key, in the same way, in the password update protocol.
\item $CS$ decrypts ($Enc_i$) and computes timestamp and examines $UP_{C_i}^{CS}$ and $MP_{C_i}^{CS}$ in the datasets. It obtains the $RR_i$ from the computation equation $RR_i= tmpRR_i\oplus N_{C_2}\oplus CSSig_1$. It extracts $PW_i$ from $tmpPW_i\oplus N_ {C_3}\oplus C_iSig_1$ and checks $PW_i$ matching in the dataset. $CS$ computes the signature ($CSSig_1 = h (TS_{C_i}\| N_{C_1}\|UP_{C_i}^{CS}\|MP_{C_i}^{CS}\|RR_i$\ $\|PW_i\|$\ $"delete")$ and then compares the result of the signatures. $CS$ computes operation $RR_i \oplus N_{C_1}$ to use $RR_i$'s signature in order to compare the $RR_i$ with the user's $R_i$. If all operations are achieved and validated correctly, $CS$ sends a request to $AS$ to check the pseudonym's association with real information and check $PW_i$. $AS$ deletes association between pseudonyms and information, and delete user's $PW_i$ from dataset. $AS$ sends MAC delete request to $CS$. Upon receiving MAC delete request, $CS$ decrypts this request, then checks security parameters and signature. If all operations are achieved and validated correctly, $CS$ deletes MAC address from dataset. At this point, this user cannot perform an authentication process in RAMHU.
\end{itemize}

\noindent Revocation from $AS$ side:
\begin{itemize}
\item $AS$ deletes the association between user's information and pseudonyms and $PW_i$ in datasets. It sends an encrypted request ($Enc_i)$ to $CS$ to delete device's MAC address for this user. After the completion of this protocol, the user is considered illegal and cannot access HC services.
\end{itemize}

\section{Security Analysis}
\label{sec:security_analysis}
In this section, a theoretical, and an experimental security analysis have been presented. We describe how RAMHU applies security requirements in protecting users' authentication information in the HC system.

\subsection{Theoretical Security Analysis}
The theoretical security analysis shows that RAMHU is secure against the various known attacks. In this section, we will show how RAMHU provides a high-security level against these attacks by providing assumptions  and proofs. Table~\ref{tab4} summarises the comparison between RAMHU and other authentication protocols in various attacks resistance.

\begin{itemize}

\item \textbf{RAMHU prevents a privileged insider attack.}\\
\textbf{Proof 1:} The internal intruder ($II$) needs to eavesdrop authentication requests between $C_i$ and $CS$. After that, he/she tries to perform an analysis of these requests based on his/her access privileges to the network. First, the analysis of these requests to obtain the private key or password is infeasible because the authentication request is encrypted by the ECIES-256 bits algorithm; $II$ cannot decrypt these requests with his private key. Second, if $II$ broke the encryption (which is impossible), he/she is unable to extract the $PW_i$ value ($tmpPW_i= PW_i\oplus N_{C_i}\oplus GM\oplus C_iSig_1$) in the login protocol because it is hidden and depends on the values of $GM$, $C_iSig_1$, and $N_ {C_i}$ where $II$ does not know the $GM$ value of the legitimate user's device, and the $C_iSig_1$ value depends on the $CM$ value that is also not known to $II$. Therefore, RAMHU prevents a privileged insider attack. 

\item \textbf{RAMHU is resistant to stolen attack.}\\
\textbf{Proof 2:} In the first case, The intruder ($I$) steals a legitimate user device (such as a laptop) that contains a client application. In our scheme, user's $PW_i$, and $IDs$ are not stored on the user's device, or in the client application. In addition, the private key is hidden and random for each authentication process by $C_iK_{pr_i}\oplus GM\oplus PW_i\oplus N_{C_i}$. Additionally, the Proof 1 shows that the $PW_i$ value cannot be extracted from the $tmpPW_i$. If $I$ is $II$, he/she also cannot use his/her $IDs$ and $PW_i$ to access information or other user data because both $CS$ and $AS$ perform matching $IDs$, and $PW_i$ to determine user-related information and data. In the second case, the attacker steals only client application. $I$ cannot use the application on another device because it does not have $OTP_i$ or original MAC, which prevents the application from being used on another device even if $I$ knows the $IDs$ and $PW_i$. The original MAC mechanism ($C_iSig_{1} = h (CM\|N_{C_{i}}\|TS_{C_{i}}$)  prevents the fake authentication request from being verified in the server because $I$ cannot generate an original MAC address ($CM$) in $C_iSig_1$. Thus, RAMHU is resistant to stolen attack.

\item \textbf{RAMHU resists replay attack.}\\
\textbf{Proof 3:} $I$ tries to get a login/registration/authentication request for a legitimate user to send it later and thus, gains access to the network. This case is infeasible in our scheme because all entities ($C_i$, $CS$, and $AS$) use timestamp (such as $TS_{CS}-TS_{C_i} \leq \triangle T$, where $\triangle T$ is the maximum transfer delay rate) that prevents the attacker from sending the authentication request at a later time. Furthermore, signatures and random nonces are not usable the next times. Hence, RAMHU successfully resists replay attack.

\item \textbf{RAMHU overcomes the MITM attack.}\\
\textbf{Proof 4:} Assume that an $I$ attempts to intercept encrypted login/authentication requests (such as $Enc_{i}(TS_{C_{i}}\|N_{C_{i}}\|$\ $CiSig_{1}\|UP^{CS}_{C_{i}}\|MP^{CS}_{C_{i}}\|tmpPW_{i}\|CiSig_{2})$) among network entities, and then modifies or replaces these requests with his/her messages to send to network entities. However, the attacker cannot replace exchanged requests among $C_i$, $CS$, and $AS$ because, first, he/she does not know the private keys ($C_iK_{pr_{i}}$, $CSK_{pr_{i}}$, $ASK_{pr_{i}}$) and therefore, the decryption process is computationally infeasible with 256 bits key length and difficulty in solving ECDLP. Second, mutual authentication with PHOTON-256 signatures prevents the modification of requests between RAMHU's entities. As a result, RAMHU gracefully overcomes the MITM attack.

\item \textbf{RAMHU is safe against guessing attack.}\\
\textbf{Proof 5:} Assume that an external intruder ($EI$) was able to penetrate the encryption ($Enc_i$) between $C_i$ and $CS$ (from Proof 4, this assumption is infeasible). This $EI$ tries to guess $PW_i$ in a login request to use it to access the network as a legitimate user. $EI$ cannot detect $PW_i$ for any authorised user (either on-line or off-line) because he/she does not know the configured process to protect $PW_i$ ($tmpPWi = PW_i\oplus N_{C_i}\oplus GM\oplus C_ {i} Sig_1$) and does not know the MAC address for that user and thus, the process of deriving $PW_i$ is infeasible (Proof 1). It is an extremely difficult process to guess $PW_i$ from the $tmpPW_i$, which is 64 hex (256 bit). In addition, $EI$ cannot detect $UID_i$ and $MID_i$ for any legitimate user because of the use of multi pseudonyms mechanism for users and medical centres instead of sending real information to legitimate users. As a result, RAMHU is safe against guessing attack.

\item \textbf{RAMHU withstands against client impersonation attack.}\\
\textbf{Proof 6:} Assume that an attacker tries to impersonate a login request  (such as $Enc_{i}$ =\ $TS_{C_{i}}\|N_{C_{i}}$\ $\|CiSig_{1}\|$\ $UP^{CS}_{C_{i}}\|MP^{CS}_{C_{i}}\|tmpPW_{i}\|CiSig_{2}$) for a legitimate user. This $I$ can create $TS_ {C_i}$ and $N_ {C_i}$ but does not know $PW_i$ and $C_iK_ {pr_i}$ (Proofs 1, and 4) for the legitimate user, namely, $I$ cannot impersonate the user identity. $I$ also tries to impersonate the legitimate user's device by programmatically changing the MAC address to a legitimate one to gain access to the network. This case is infeasible because the original MAC check ($CM$) in $CiSig_{1} = h (CM\|N_{C_{i}}\|TS_{C_{i}})$ detects the attacker's attempt to mimic the legitimate user's device (as in Proof 2). Therefore, RAMHU withstands against instances of impersonating the user's identity and device.

\item \textbf{RAMHU resists server impersonation attack.}\\
\textbf{Proof 7:} Assume that an $I$ traps login requests from $C_i$ to $CS$. The attacker tries to deceive $C_i$ by sending fake requests to $C_i$ in order to inform them that he is a legitimate server. $I$ needs the private key for CS to decrypt and to accomplish the attack. Mutual authentication prevents $I$ from impersonating $CS$'s requests (such as $Enc_i(TS_{CS}\|N_{CS}\|UP_{CS}^{C_i}\|MP_{CS}^{C_i}\|CSSig_5)$) and sending them to $C_i$. This mechanism ensures that $C_i$ deals with legitimate $CS$. Consequently, our protocol effectively resists server impersonation attack.

\item \textbf{RAMHU resists DoS attack.}\\
\textbf{Proof 8:} In order for $I$ to execute a DoS attack against $CS$ and $AS$, he/she needs to decrypt the login request and change its data or send the same request multiple times for destroying servers. However, in the first case, decryption and change of signatures are infeasible as in Proofs 2 and 4. $CS$ checks signature validity and rejects login requests containing fake signatures, and $I$ cannot execute a collision or preimage attack because PHOTON-256 supplies PR, SPR, and CR. In the second case, the attacker sends the same request multiple times. This status is infeasible because CS or AS checks timestamp ($TS_{C_i}$, $TS_{CS}$, $TS_{AS}$) for each login/authentication request and eliminates all late requests (Proof 3) without checking the other security parameters such as $PW_i$, $CM$, and multi pseudonyms. In case $I$ can break the encryption, he/she can change timestamp and nonce but cannot tamper with the signatures. RAMHU prevents this condition during $C_iSig_1$ and does not need to check the remaining security parameters. Therefore, RAMHU successfully resists DoS attack.

\item \textbf{RAMHU is secure against password change attack.}\\
\textbf{Proof 9:} Assume that an $I$ intercepts a request to change $PW_i$ between $C_i$ and $CS$. When $I$ obtains an encrypted password update request (such as $Enc_i (TS_{C_{i}}\|NC_1\|NC_2\|NC_3\|UP_{C_{i}}^{CS}\|$\ $MP_{C_{i}}^{CS}\|$ $tmp\_oldPW_i\|$\ $tmp\_newPW_i\|$\ $C_{i}Sig_{1})$). If $I$ can decrypt (this process is infeasible as in Proofs 1, and 4), he/she will find a temporary passwords and cannot derive new $PW_i$ because it depends on the $C_iSig_1 = h (TS_{C_{i}}\|NC_1\|UP_{C_{i}}^{CS}\|$\ $MP_{C_{i}}^{CS}\|old PW_i)$. The signature operation ($C_iSig_1$) is based on the old $PW_i$ which is not explicitly sent to $CS$ in this protocol. $I$ does not know old $PW_i$ and therefore, he/she cannot create a signature to complete the $PW_i$ change process. Thereupon, RAMHU provides a reliable solution against password change attack.

\item \textbf{RAMHU is resistant to eavesdropping attack.}\\
\textbf{Proof 10:} Assume that an $I$ eavesdrops login/authentication requests to gain information about user authentication and access to the network. However, in our scheme, $I$ will not benefit from requests that are intercepted because RAMHU uses ECIES algorithm with key 256 bits to encrypt authentication information. The attacker can only decrypt requests by deriving private keys and this operation is infeasible (as in Proofs 1, and 2) due to key length, and random encryption with nonces (anonymity). Therefore, our protocol is resistant to eavesdropping attack.

\item \textbf{RAMHU resists traceability attack.}\\
\textbf{Proof 11:} $I$ attempts to collect as many login/authentication requests as possible and then performs an analysis of those requests that helps him/her to perform user identity tracing. When an $I$ succeeds in tracking user requests, he/she can detect and distinguish patients' data. All exchanged requests among RAMHU's entities do not contain direct users' information (such as username). RAMHU replaces the real user $ID_s$ ($UID_i$ and $MID_i$) with pseudonyms. Our protocol uses multi pseudonyms ($UP_{C_i}^{CS}$, $UP_{CS}^{AS}$, $UP_{AS}^{CS}$, and $UP_{CS}^{C_i}$ for users and $MP_{C_i}^{CS}$, $MP_{CS}^{AS}$, $MP_{AS}^{CS}$, and $MP_{CS}^{C_i}$ for medical centres) to prevent attackers from tracking user requests and revealing their identities when transferred between RAMHU entities ($C_i$, $CS$, and $AS$). Hence, RAMHU resists traceability attack.

\item \textbf{RAMHU prevents revocation attack.}\\
\textbf{Proof 12:} Assume that an $I$ tries to penetrate a revocation request. The attacker tries to analyse the request and use it to prevent users from accessing the network's services. Depending on Proofs 1, 5, and 11, the attacker cannot extract or distinguish $UID_i$, $MID_i$ and $PW_i$ from the revocation request. The attacker does not know and cannot extract a reason from the $tmpRR_i$, which is based on the values of $RR_i$, $N_{C_2}$, and $C_iSig_1$. In Proofs 2 and 8, $I$ cannot perform collision, preimage, and second preimage attacks against the PHOTON-256 algorithm. Thus, our protocol prevents penetration of revocation request.

\item \textbf{RAMHU resists verifier attack.}\\
\textbf{Proof 13:} Assume that $I$ tries to penetrate the datasets in $CS$. If the attacker is $EI$, he/she cannot penetrate datasets because he/she does not have a $K_{pr}$, $UID$, $MID$, $OTP$, and $PW_i$. If the attacker is $II$, when he penetrates datasets in $CS$ and wants to impersonate another user's identity, first, he/she cannot distinguish this information for a particular user because the real information for users is stored in $AS$. Second, since $CS$ does not contain passwords' dataset, $II$ will not benefit from hacking datasets such as pseudonyms and cannot create a request and send it to $AS$ because it does not know users' passwords. Therefore, RAMHU resists verifier attack meritoriously.

\item \textbf{RAMHU resists leakage attack.}\\
\textbf{Proof 14:} Suppose that $I$ listens to some exchanged requests among $C_i$, $CS$, and $AS$ and tries to find any information that helps him/her to penetrate network authentication such as sending an ID explicitly, or sending a password with weak encryption. Exchanged requests between RAMHU entities such as a password update request ($Enc_i(TS_{C_i}\|N_{C_1}\|N_{C_2}$\ $\|N_{C_3}$\ $\|UP_{C_i}^{CS}\|MP_{C_i}^{CS}\|$\ $tmp\_oldPW_i\|$\ $tmp\_newPW_i\|C_iSig_1)$) show that $I$ does not receive any leaked real information for users during transmission, such as $IDs$ and $PW_i$ (all information is anonymous and hidden) that could be useful in penetrating the HC network. Therefore, RAMHU resists leakage attack.

\end{itemize}

\begin{table}[!t]
\begin{center} 
\caption{Comparison of resistance in repelling the threats between RAMHU and existing schemes} 
\label{tab4}
\scriptsize
\setlength{\tabcolsep}{2pt}
\begin{tabular}{|p{95pt}|p{25pt}|p{25pt}|p{25pt}|p{25pt}|p{25pt}|p{25pt}|p{25pt}|p{25pt}|p{45pt}|p{32pt}|}
\hline
Attack  &He et al. \cite{tp24} &Giri et al. \cite{tp28} &Li et al. \cite{tp26} &Farash et al. \cite{tp23} &Kumar et al. \cite{tp16} &Jiang et al. \cite{tp14} &Rajput et al. \cite{tp37} &Das et al. \cite{tp15} &Chandrakar et al.               \cite{tp50} & RAMHU scheme\\
\hline
Privileged-insider                &                     &\checkmark &\checkmark &                     &                     &\checkmark &                    &\checkmark &\checkmark &\checkmark\\
Stolen device/application  &                     &                     &\checkmark &                     &                     &\checkmark &                    &\checkmark &\checkmark &\checkmark\\
Replay	                               &\checkmark &\checkmark &\checkmark &\checkmark &\checkmark &                     &\checkmark &\checkmark &\checkmark &\checkmark\\
MITM                                   &\checkmark &                     &                     &\checkmark &                     &\checkmark &                     &\checkmark &                     &\checkmark\\
Guessing                              &                     &\checkmark &\checkmark &                     &                      &\checkmark &                    &\checkmark &\checkmark &\checkmark\\
Client impersonation          &\checkmark &\checkmark &\checkmark &\checkmark &\checkmark  &\checkmark &\checkmark &\checkmark &\checkmark &\checkmark\\
Server impersonation         &\checkmark &\checkmark &\checkmark &\checkmark &\checkmark  &\checkmark &                     &\checkmark &\checkmark &\checkmark\\
DoS                                       &                     &\checkmark &\checkmark &                      &                     &                     &\checkmark &                     &                     &\checkmark\\
Change password              &                      &\checkmark &\checkmark &                      &                     &\checkmark &                     &\checkmark &\checkmark &\checkmark\\
Eavesdropping                  &\checkmark   &\checkmark &\checkmark & \checkmark &\checkmark &\checkmark &\checkmark &\checkmark &\checkmark &\checkmark\\
Traceability                        &\checkmark   &                     &\checkmark & \checkmark &\checkmark &\checkmark &\checkmark &                      &\checkmark &\checkmark\\
Revocation                        &                        &                    &                     &                       &                     &\checkmark & \checkmark &                     &                     &\checkmark\\
Verifier                                &\checkmark    &                    &\checkmark &                       &                     &\checkmark &                      &\checkmark &                    &\checkmark\\
Leakage                              &\checkmark   &                    &                     &\checkmark   &                      &\checkmark &\checkmark  &\checkmark &\checkmark &\checkmark\\
\hline
\end{tabular}
\end{center}
\end{table}

\subsection{Experimental Security Analysis}
To ensure that authentication schemes work correctly and are authenticated, these schemes require a formal tool to detect the robustness of users' authentication in the network. We provide the proposed scheme simulation using the AVISPA tool to verify our scheme, whether safe or unsafe. Our simulations are based on the AVISPA tool with the current version v.1.1 (13/02/2006) available on the website in \cite{tp75}. This tool has been widely used and accepted by researchers in recent years \cite{tp72,tp73}. It has been used to check security problems and ensure that known attacks are not able to penetrate users' authentication information.

\subsubsection{AVISPA Description}
After designing any authentication protocol, this protocol should be checked and its accuracy verified under a test model such as the Dolev-Yao (dy) to analyse, trace and detect the possibility of attack theoretically. However, this analysis needs to be simulated in a practical manner to detect errors and hidden traces of the authentication protocol designer, statistics and accurate results, checking several techniques on the same protocol. The AVISPA tool provides the features listed above as well as the ease, robustness, and applicability of this tool to implement authentication protocols \cite{tp74}.\\

The AVISPA tool is a push-button, testing/proofing model and is based on HLPSL language. This language uses directives and expressive terms to represent security procedures. It is integrated with four backends that are the On-the-Fly Model-Checker (OFMC), the Constraint-Logic-based Attack Searcher (CL-AtSe), the SAT-based Model-Checker (SATMC), and (Tree Automata based on Automatic Approximations for the Analysis of Security Protocols (TA4SP) to perform the simulation in AVISPA. Each backend gives the result of simulation analysis statistics that is different from the other \cite{tp75}. SATMC and TA4SP backends do not work with a security protocol that implements the XoR gateway; therefore, we relied on OFMC and CL-AtSe backends to simulate RAMHU. To implement the authentication protocol in AVISPA, this protocol should be first written in HLPSL and then converts to the low-level language. The latter is read directly by the backends, which is an intermediate format (IF) by the hlpsl2if compiler. Then, it converts to an output format (OF) to extract and describe the result of analysis by one of four backends. The result of the analysis accurately describes that the protocol is safe or not safe with some statistical numbers. HLPSL is modular, and role-oriented. This language allows the completion of authentication protocol procedures as well as intruder actions. To represent authentication scenarios and build simulation structures, HLPSL uses roles, including basic roles such as clients and servers (clients, centralServer, and attributesServer), and composition (session and environment) roles that control the sequence of sending and receiving actions between clients and servers. In addition, communication channels between network entities are governed by the dy model. Many basic types used in HLPSL to represent variables, constants, and algorithms (symmetric/asymmetric/hash) such as agent, public\_key, message, text, nat, const, and hash\_func; in addition to some symbols and terms that have shown in Table~\ref{tab5}, more details are provided in \cite{tp75}. The authentication protocol in AVISPA depends on security features in the goal specification. Each protocol contains a set of goals (authentication and secret) in authenticating each party with the other. The goals in secrecy\_of demonstrate that secrets are not exposed or hacked to non-intended entities. While the goals in authentication\_on demonstrate that strong authentication has been applied between entities based on witness and request.

\begin{table}[!t]
\begin{center} 
\caption{Some HLSPL's symbols and statements} 
\label{tab5}
\scriptsize
\setlength{\tabcolsep}{2pt}
\begin{tabular}{|p{95pt}|p{230pt}|}
    \hline
    Symbol   	                     & Description\\ 
    \hline
     .                               & Concatenation\\ 			
    $\{\}\_Kp$                       & Asymmetric encryption with public key\\			
	$played\_by$	                 & Used to link the role with the intended entity\\
    $=|>$                            & Reaction transitions to relate event with act\\ 
    /\textbackslash                  & Conjunction\\
    $protocol\_id$                   & Goal identifier\\
    $secrecy\_of$                    & The goal of protecting the secret between entities permanently\\
    $authentication\_on$             & The goal of strong authentication between entities\\ 
    $intruder\_knowledge$            & What intruder knows about network\\
                    
    \hline
\end{tabular}
\end{center}
\end{table}

\subsubsection{Proposed Scheme with AVISPA}
In this section, we will illustrate the simulation of RAMHU in the AVISPA tool using HLPSL language. Our scheme depends on three core roles: clienti, centralServer, and attributesServer played by $C_i$, $CS$ and $AS$ respectively. In addition, it includes the supporting roles such as session, and environment, goal specification section. Each role contains parameters, variables, and local constants. Each basic role contains a transition section that indicates the sequence of communication among entities. Each supporting role contains a composition section that indicates the binding of roles and sessions. Asymmetric encryption has been implemented between scheme entities ($C_i$, $CS$, and $AS$) during public key exchange ($KC_{pu}$, $KCS_{pu}$, and $KAS_{pu}$) to perform confidentiality. Moreover, mutual authentication has used to ensure the legitimacy of related parties in the protocols of the proposed scheme (initial setup, registration, login, and authentication). Moreover, it uses nonces ($N_c$, $N_{cs}$, and $N_{as}$) and timestamp ($TS_c$, $TS_{cs}$, $TS_{as}$) to support the features of anonymity and freshness. Our scheme accomplishes 10 secrecy goals and 6 authentication goals as noted in the goal section in Figure~\ref{fig12}.\\

The authentication process begins by sending requests from clients to the server. Therefore, the $client_i$ role includes the start signal as shown in Figure~\ref{fig9}. $C_i$ receives the start signal and changes the state flag (State variable) from 0 to 1. It replaces $UID$ and $MID$ with $UP_c$ and $MP_c$ and calculates the timestamp ($TS_c'$) and new nonce ($N_c'$) by the new() operation. After that, it computes the signatures ($C_iSig_1$ and $C_iSig_2$), as well as the password hiding process as calculated in the computation operation of the $TmpPW$ parameter. It encrypts the registration and login request ($TS_c'$, $N_c'$, $GM'$, $C_iSig_1'$, $UP_c'$, $MP_c'$, $OTP_i$, $Tmp\_PW'$, and $C_iSig_2'$) by the public key ($KCS_{pu}$) to establish a reliable communication with $CS$. $C_i$ sends the request to $CS$ where the transmission process is performed by the SND() operation. It achieves a set of secret goals ($sec1$ to $sec6$) with both $CS$ and $AS$; these secrets have been only known and kept by the intended parties. For instance in $sec1$, $UID$, $MID$, and $PW$ have been known and kept only to $C_i$, and $AS$. While in $sec3$ $GM$, and $CM$ have been known and kept only to $C_i$ and $CS$, since these parameters are not transmitted directly during the transition of information between network parties such as $PW$ implicitly is in the computation of $TmpPW$ and $CM$ is implicitly in $C_iSig_1$. $C_i$ also achieves the goal of authentication using statement (witness) with parameters ($C_i$, $CS$, $ci\_cs\_auth2$, {$N_{cs}$, $TS_c$}). Namely, $C_i$ is a witness that the security parameters ($N_{cs}$, $TS_c$) are fresh and correct. $CS$ uses a statement (request) to validate parameters with the strong authentication goal ($ci\_cs\_auth2$) specified in the goal section (Figure 12). $C_i$ receives the authentication response by the RCV () operation and sent from $AS$ by $CS$. Then, $C_i$ decrypts the response using its private key to verify the security parameters. If all security parameters are verified correctly, $C_i$ performs the mutual authentication process securely.
\begin{figure}[!ht]
   \begin{minipage}[t]{0.48\textwidth}
     \scriptsize
       \begin{SaveVerbatim}{VerbCode}
role clienti(Ci,CS,AS:agent,  KCpu,KCSpu:public_key,
                   H:hash_func, 
                   UID,MID,OTPi,PW,GM,CM:message,
                   SND,RCV:channel(dy)) 
played_by Ci def=
local 
     State:nat, 
     TSc,TScs,Nc,Ncs:text,
     CiSig1,CiSig2:text,
     UPc,MPc,UPcs,MPcs,TmpPW:message
const
     sec1,sec2,sec3,sec4,sec5,sec6,
     cs_ci_auth1,ci_cs_auth2:protocol_id
init 
     State := 0
transition  
1. State  = 0 /\ RCV(start) =|> 
   State':= 1 
   /\ UPc':=UID/\MPc':=MID/\GM':=Ci
   /\ Nc':=new()/\ TSc':=new()
   /\ CiSig1':= H(CM.Nc'.TSc')
   /\ CiSig2':= H(GM'.Nc'.TSc'.CiSig1'.UPc'.MPc'.
                  OTPi.PW)
   /\ TmpPW':=xor(PW,xor(Nc',xor(GM',CiSig1')))

% registeration and login protocol
% Ci sends security parameters to CS 
   /\ SND({TSc'.Nc'.GM'.CiSig1'.UPc'.MPc'.OTPi.
           TmpPW'.CiSig2'}_KCSpu) 
   /\ secret({UID,MID,PW},sec1,{Ci,AS})
   /\ secret(PW,sec2,{Ci,CS})
   /\ secret({GM,CM},sec3,{Ci,CS})
   /\ secret({CiSig1',CiSig2'},sec4,{Ci,CS})
   /\ secret({TmpPW',OTPi},sec5,{Ci,CS})
   /\ secret({UPc,MPc,UPcs,MPcs},sec6,{Ci,CS})
		   
2. State  = 1 
% Ci receives authentication response from CS
   /\ RCV({TScs'.Ncs'.Nc.UPcs'.MPcs'. 
           H(TScs'.Ncs'.UPcs'.MPcs')}_KCpu)=|> 
   State':= 2 /\ SND({Ncs'}_KCSpu) 

% Clienti checks that the received security parameters
   /\ request(Ci,CS,cs_ci_auth1,{Nc,TScs})
   /\ request(Ci,CS,cs_ci_auth5,{TmpPW,OTPi})
% Clienti sends Ncs to prove her identity
   /\ witness(Ci,CS,ci_cs_auth2,{Ncs,TSc})
end role
   \end{SaveVerbatim}
  \setlength{\fboxsep}{1mm}  
  \fbox{\BUseVerbatim{VerbCode}}
     \caption{Client role in HLPSL}\label{fig9}
   \end{minipage}\hfill
   \begin {minipage}[t]{0.48\textwidth}
     \centering
     \scriptsize
     \begin{SaveVerbatim}{VerbCode}
role centralServer(Ci,CS,AS:agent, 
                   KCpu,KCSpu,KASpu:public_key,
                   H:hash_func, 
                   SND,RCV:channel(dy)) 
played_by CS def=
local 
     State:nat,
     UPc,MPc,UPcs,MPcs,UPas,MPas,TmpPW:message,
     CSSig1,CSSig2,CSSig3:text,
     TSc,TScs,TSas,Nc,Ncs,Nas:text,
     PW,OTPi:message,
     GM,CM:message
const
     sec1,sec2,sec3,sec4,sec5,sec6,sec7,sec8,sec9,sec10,
     ci_cs_auth2,cs_ci_auth1,cs_ci_auth5,cs_as_auth3,
     as_cs_auth4,as_cs_auth6:protocol_id
init 
     State := 0
transition
% Registration and login request from Ci 
1. State  = 0 /\ RCV({TSc'.Nc'.GM'.H(CM'.Nc'.TSc').
                      UPc'.MPc'.OTPi'.TmpPW'.
                      H(GM'.Nc'.TSc'.H(CM'.Nc'.TSc').
                      UPc'.MPc'.OTPi'.PW')}_KCSpu) =|> 
   CSSig1':= H(CM'.Nc'.TSc')
   /\ CSSig2':= H(GM'.Nc'.TSc'.CSSig1'.UPc'.MPc'.OTPi.PW)
   /\ PW':=xor(TmpPW,xor(Nc',xor(GM',CSSig1)))          
   /\ secret(PW',sec2,{CS,Ci})
   /\ secret({GM,CM},sec3,{CS,Ci})
   /\ secret({CSSig1,CSSig2},sec4,{CS,Ci})
   /\ secret({TmpPW',OTPi},sec5,{CS,Ci})
   /\ secret({UPc,MPc,UPcs,MPcs},sec6,{CS,Ci})                      

% Authentication request from CS to AS
   /\ State':= 1 /\Ncs':=new() /\TScs':=new()
   /\ CSSig3':= H(TScs'.Ncs'.UPcs.MPcs)
   /\ TmpPW':=xor(PW,xor(Ncs',CSSig3'))
   /\ SND({TScs'.Ncs'.UPcs.MPcs.CSSig3'.TmpPW'}_KASpu)
   /\ secret(PW,sec7,{CS,AS})
   /\ secret({CSSig3},sec8,{CS,AS})
   /\ secret({TmpPW'},sec9,{CS,AS})
   /\ secret({UPcs,MPcs,UPas,MPas},sec10,{CS,AS}) 

% Authentication response from AS to CS
2. State = 1 /\ RCV({TSas'.Nas'.Ncs.UPas'.MPas'.
                     H(TSas'.Nas'.UPas'.MPas')}_KCSpu)=|>
% Authentication response to Ci
   State':= 2/\ Ncs':=new() /\TScs':=new()
   /\ SND({TScs'.Ncs'.Nc.UPcs.MPcs.
           H(TScs'.Ncs'.UPcs.MPcs)}_KCpu)
   % CS prove his identity
   /\ witness(CS,Ci,cs_ci_auth1,{Nc,TScs'})
   /\ witness(CS,Ci,cs_ci_auth5,{TmpPW,OTPi})
   /\ witness(CS,AS,cs_as_auth3,{Nas',TScs'})
   /\ request(CS,AS,as_cs_auth4,{Ncs',TSas})
   /\ request(CS,AS,as_cs_auth6,{PW})   
3. State  = 2 /\ RCV({Ncs}_KCSpu) =|> 
% CS checks that the received nonce and timestamp correct
   State':= 3 /\ request(CS,Ci,ci_cs_auth2,{Ncs,TSc})
end role
   \end{SaveVerbatim}
  \setlength{\fboxsep}{1mm}  
  \fbox{\BUseVerbatim{VerbCode}}
     \caption{Central server role in HLPSL}\label{fig10}
   \end{minipage}
\end{figure}

\begin{figure}[!ht]
   \begin{minipage}[t]{0.48\textwidth}
     \scriptsize
       \begin{SaveVerbatim}{VerbCode}
role attributesServer(AS,Ci,CS:agent,
                      KASpu,KCSpu:public_key,
                      H:hash_func,SND,RCV:channel(dy))
played_by AS def=
local
     State:nat,
     UID,MID,PW:message,
     TmpPW:message,
     ASSig1,ASSig2:text,
     TScs,Ncs,TSas,Nas:text,
     UPcs,MPcs,UPas,MPas:message,
     GM,CM:message
const
     sec1,sec7,sec8,sec9,sec10,
     cs_as_auth3,as_cs_auth4,as_cs_auth6:protocol_id
init
     State:= 0
transition
1. State = 0 /\ RCV({TScs'.Ncs'.UPcs'.MPcs'.
                     H(TScs'.Ncs'.UPcs'.MPcs').
                     TmpPW'}_KCSpu)=|>
   ASSig1':= H(TScs'.Ncs'.UPcs.MPcs)
   /\ PW':=xor(TmpPW,xor(Ncs',ASSig1))
   /\ secret({UID,MID,PW},sec1,{AS,Ci})
   /\ secret(PW,sec7,{AS,CS})
   /\ secret({ASSig1},sec8,{AS,CS})
   /\ secret({TmpPW'},sec9,{AS,CS})
   /\ secret({UPcs,MPcs,UPas,MPas},sec10,{AS,CS}) 
   /\ State':= 1 
   /\ Nas':=new() 
   /\ TSas':=new()
   /\ ASSig2':=H(TSas'.Nas'.UPas.MPas)
   /\ SND({TSas'.Nas'.Ncs.UPas.MPas.ASSig2'}_KCSpu)  
   /\ witness(AS,CS,as_cs_auth4,{Ncs',TSas'}) 
   /\ witness(AS,CS,as_cs_auth6,{PW'})
   /\request(AS,CS,cs_as_auth3,{Nas',TScs'})
end role
   \end{SaveVerbatim}
  \setlength{\fboxsep}{1mm}  
  \fbox{\BUseVerbatim{VerbCode}}
     \caption{Attribute server role in HLPSL}\label{fig11}
   \end{minipage}\hfill
   \begin {minipage}[t]{0.48\textwidth}
     \centering
     \scriptsize
     \begin{SaveVerbatim}{VerbCode}
role session(Ci,CS,AS:agent, 
             KCpu,KCSpu,KASpu:public_key, 
             H:hash_func,
             UID,MID,OTPi,PW,GM,CM:message) 
def=
local 
     SndC,RcvC,SndCS,RcvCS,SndAS,RcvAS:channel(dy)
composition 
     clienti(Ci,CS,AS,KCpu,KCSpu,H,UID,
             MID,OTPi,PW,GM,CM,SndC,RcvC)
     /\ centralServer(Ci,CS,AS,KCpu,KCSpu,
                      KASpu,H,SndCS,RcvCS)
     /\ attributesServer(AS,Ci,CS,KASpu,KCSpu,
                         H,SndAS,RcvAS)
end role
%%%
role environment() 
def=
const 
     ci,cs,as,i_ci,i_cs,i_as:agent,
	 kCpu,kCSpu,kASpu,ki:public_key,
     uid,mid,otp,pw,gm,cm:message,
     h:hash_func,
     sec1,sec2,sec3,sec4,sec5,
     sec6,sec7,sec8,sec9,sec10,
     ci_cs_auth2,cs_ci_auth1,cs_ci_auth5,cs_as_auth3,
     as_cs_auth4,as_cs_auth6:protocol_id
     intruder_knowledge={ci,cs,as,i_ci,i_cs,i_as,
                         kCpu,kCSpu,kASpu,ki}
composition
     session(ci,cs,as,kCpu,kCSpu,kASpu,h
	         ,uid,mid,otp,pw,gm,cm)
     % Check replay attack
     /\ session(ci,cs,as,kCpu,kCSpu,kASpu,h
                ,uid,mid,otp,pw,gm,cm) 
     % Check MITM attack
     /\ session(cs,ci,as,kCSpu,kCpu,kASpu,h
                ,uid,mid,otp,pw,gm,cm) 
     % Check impersonate Ci
     /\ session(i,cs,as,ki,kCSpu,kASpu,h
                ,uid,mid,otp,pw,gm,cm)  
     % Chekc impersonate CS  
     /\ session(ci,i,as,kCpu,ki,kASpu,h
                ,uid,mid,otp,pw,gm,cm) 
     % Check impersonate AS    
     /\ session(ci,cs,i,kCpu,kCSpu,ki,h
                ,uid,mid,otp,pw,gm,cm)      
end role
%%%
goal
 secrecy_of sec1,sec2,sec3,sec4,sec5,
            sec6,sec7,sec8,sec9,sec10
 authentication_on cs_ci_auth1,ci_cs_auth2
                  ,cs_as_auth3,as_cs_auth4,
                  cs_ci_auth5,as_cs_auth6
end goal
   \end{SaveVerbatim}
  \setlength{\fboxsep}{1mm}  
  \fbox{\BUseVerbatim{VerbCode}}
     \caption{Supporting roles in HLPSL}\label{fig12}
   \end{minipage}
\end{figure}
As shown in Figure~\ref{fig10}, $CS$ receives a registration and login request by RCV () operation in state 0. It decrypts request with its private key and then checks the parameters and signatures to accomplish secret and authentication goals. CS changes the state signal from 0 to 1 and constructs an authentication request based on the security parameters ($TS_{cs}'$, $N_{cs}'$, $UP_{cs}$, $MP_{cs}$, $CSSig_3$, and $TmpPW$) and encrypts it by the public key of $AS$. $CS$ performs strong authentication with $AS$ during witness ($CS$, $AS$, $as\_cs\_auth4$, {$TS_{cs}'$, $N_{cs}'$}) to accomplish the authentication goal ($as\_cs\_auth4$) based on timestamp and fresh nonce and validated in $AS$ by statement (request). In state 1, $CS$ receives an authentication response from $AS$ and checks the security parameters after decryption with its private key. It changes the state signal to 2 and then constructs the authentication response to $C_i$. $CS$ sends response with two strong authentication goals ($cs\_ci\_auth1$ and $cs\_ci\_auth5$) based on the security parameters ($N_c$, $TS_{cs}$, $TmpPW$, and $OTP_i$).\\

$AS$ receives the authentication request and decrypts it with the private key as shown in Figure~\ref{fig11}. It accomplishes 5 secret goals, and accomplishes two authentication goals ($as\_cs\_auth4$ and $as\_cs\_auth6$) based on $TS_{cs}'$, $N_{cs}'$, and $PW '$. It constructs the authentication response and establishes strong authentication when validating parameters ($TS_{as}$, $N_{cs}'$, and $PW'$) in $CS$. Figure~\ref{fig12} displays the roles of session, environment, and goal section. In the session role, a composition process has been performed for the three roles ($clienti$, $centralServer$, and $attributeServer$). This role specifies the transmit and receive channels in the dy model. In the environment role, the security parameters, the goals specified in the goal section, and the known information for the intruder ($intruder\_knowledge$) have been defined. In this role, one or more sessions are composed. We tested our scheme with sessions for replay, MITM, and impersonating attacks. We assumed that an intruder ($I$) creates a public key ($ki$) and has knowledge of the public keys ($kCpu$, $kCSpu$, and $kASpu$) of legitimate entities in the network. The intruder attempts to resend the registration/login or authentication requests later, intercept/modify these requests, or impersonate the participating entities using $i\_ci$, $i\_cs$, and $i\_as$ constants rather than $ci$, $cs$, and $as$. The results section shows that these attacks cannot penetrate the security goals in our scheme.

\subsubsection{Simulation Results}
In this section, the simulation results in the AVISPA tool are based on two backends (OFMC, and CL-AtSe). Figure~\ref{fig13} shows the simulation result with the OFMC backend and Figure~\ref{fig14} displays the simulation result with the CL-AtSe backend. From the results shown in Figure~\ref{fig13} and Figure~\ref{fig14}, our scheme clearly and accurately shows the SAFE result in the SUMMARY section, bounded number of sessions in the DETAILS section, the goals of the scheme achieved (as\_specified) in the GOAL section as well as statistical numbers such as time, number of nodes, and analysed states in the STATISTICS section for both figures. Based on these results, we note that our scheme is capable of preventing passive and active attacks such as replay, MITM, and impersonating. Thus, the goals of the scheme in Figure~\ref{fig12} achieved to prevent the violation of legitimate user information in the network authentication.

\begin{figure}[!ht]
   \begin{minipage}[t]{0.48\textwidth}
     \scriptsize
       \begin{SaveVerbatim}{VerbCode}
% OFMC
% Version of 2006/02/13
SUMMARY
 SAFE
DETAILS
 BOUNDED_NUMBER_OF_SESSIONS
PROTOCOL
 /home/span/span/testsuite/results/RAMHU.if
GOAL
 as_specified 
BACKEND
 OFMC
COMMENTS
STATISTICS
 parseTime: 0.00s
 searchTime: 54.60s
 visitedNodes: 3636 nodes
 depth: 11 plies
   \end{SaveVerbatim}
  \setlength{\fboxsep}{1mm}  
  \fbox{\BUseVerbatim{VerbCode}}
     \caption{Simulation result using OFMC}\label{fig13}
   \end{minipage}\hfill
   \begin {minipage}[t]{0.48\textwidth}
     \centering
     \scriptsize
     \begin{SaveVerbatim}{VerbCode}
SUMMARY
  SAFE

DETAILS
  BOUNDED_NUMBER_OF_SESSIONS
  TYPED_MODEL

PROTOCOL
  /home/span/span/testsuite/results/RAMHU.if

GOAL
  As Specified

BACKEND
  CL-AtSe

STATISTICS

  Analysed   : 324 states
  Reachable  : 64 states
  Translation: 0.52 seconds
  Computation: 0.42 seconds
   \end{SaveVerbatim}
  \setlength{\fboxsep}{1mm}  
  \fbox{\BUseVerbatim{VerbCode}}

     \caption{Simulation result using CL-AtSe}\label{fig14}
   \end{minipage}
\end{figure}

\section{Conclusion and Future Work}
\label{sec:conclustion}
In this paper, we found that existing healthcare applications were vulnerable to weak security against some known attacks. Towards this end, we proposed a new robust authentication protocol (RAMHU) to prevent internal, external, passive, and active attacks. Our scheme uses multi pseudonyms for both users and medical centres to prevent the transmission of real information in the authentication request and MAC address to prevent counterfeit devices from connecting to the network. The lightweight encryption and signature algorithms (as described in Section 3) is used to ensure that RAMHU's efficient interaction with user requests is ensured. In addition, we provided a formal and informal security analysis to demonstrate the effectiveness of RAMHU in repelling known attacks. The RAMHU scheme provides a high-level security that maintains authentication information for users against various attacks. Future directions for further the development of this scheme are as follows:
\begin{enumerate}
\item RAMHU requires strong authorization to support patient data exchange after user authentication. We need a mechanism that supports role-based and attribute-based privileges (e.g., doctor, nurse, patient, advisor, researcher, and emergency) to access patients' data on a data server ($DS$). We intend to use authorization policies with signatures to ensure proper authorization of access to the data repository.
\item Our scheme uses lightweight and efficient performance algorithms that, according to many researchers, have shown that ECIES and PHOTON are efficient encryption and signature algorithms, respectively. We intend to evaluate RAMHU in terms of efficiency and discovery of performance standards such as end-to-end request delay, throughput, and error rate.
\item We intend to use the wireless sensor network (WSN) to collect patients' data and send it to $DS$. However, collecting and storing data in $DS$ requires the use of encryption and signature algorithms against known attacks. 
\end{enumerate}

\section*{Data Availability}
 No data were used to support this study. 
\section*{Conflicts of Interest}
The authors declare that they have no conflicts of interest.
\section*{Acknowledgments}
This research did not receive specific funding but was performed as part of the employment of the authors.

\bibliographystyle{IEEEtranN}
\bibliography{ref}
\end{document}